\documentclass[twocolumn,aps,prb,superscriptaddress]{revtex4-2}
\usepackage{color}
\usepackage{mathrsfs}
\usepackage{bm}
\usepackage{amsmath}
\usepackage{amssymb}
\usepackage{graphicx}
\usepackage{hyperref}
\usepackage{color}
\usepackage{siunitx}
\usepackage{braket}
\usepackage{tikz}
\usepackage{multirow}
\usepackage{array}
\allowdisplaybreaks[1]

\makeatletter
\hypersetup{hypertex=true, colorlinks=true, linkcolor=blue, anchorcolor=blue,citecolor=blue}

\usepackage{dcolumn}
\usepackage{bm}
\usepackage{mathrsfs}
\usepackage{amsfonts}
\usepackage{xcolor}
\usepackage{epstopdf}

\begin{document}
\title{Algebraic Criterion and Graph-Theoretic Construction of Intrinsic Superconducting Diode Effects}
\author{Ran Wang}
\email{sa20168043@mail.ustc.edu.cn}
\affiliation{Anhui Provincial Key Laboratory of Low-Energy Quantum Materials and
Devices, High Magnetic Field Laboratory, HFIPS, Chinese Academy
of Sciences, Hefei, Anhui 230031, China}
\affiliation{Science Island Branch of Graduate School, University of Science and
Technology of China, Hefei, Anhui 230026, China}
\author{Ning Hao}
\email{haon@hmfl.ac.cn}
\affiliation{Anhui Provincial Key Laboratory of Low-Energy Quantum Materials and
Devices, High Magnetic Field Laboratory, HFIPS, Chinese Academy
of Sciences, Hefei, Anhui 230031, China}
\affiliation{Science Island Branch of Graduate School, University of Science and
Technology of China, Hefei, Anhui 230026, China}

\begin{abstract}
The intrinsic superconducting diode effect (SDE) is distinguished from the Josephson diode effect (JDE) by its manifestation of nonreciprocal critical current phenomena within a monolithic superconductor, typically linked to finite-momentum Cooper pairing. The long-standing assumption that SDE requires co-breaking of time-reversal and inversion symmetries proves to be necessary but not sufficient. In this work, we propose an algebraic diagnostic criterion for intrinsic SDE, expressed as two inequalities evaluated directly from the bare Hamiltonian. This criterion further reveals a graph-theoretic construction for nonreciprocal models, offering design principles that extend beyond superconductivity.
\end{abstract}
\maketitle

\section{Introduction}
\begin{figure}[htp]
	\centering
    \includegraphics[width=1.0\columnwidth]{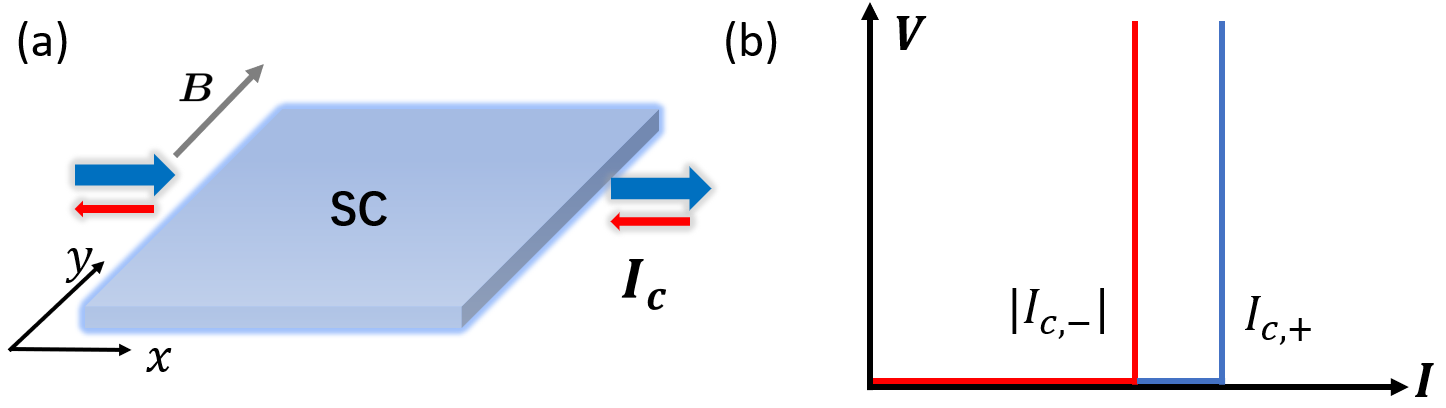}
	\caption{Schematic of SDE in the helical superconductors. (a) SDE is achieved in these two dimensional systems, when an in-plane magnetic field is applied perpendicular to the current. (b) The curvatures of $I-V$, in which the difference between the forward $I_{c,+}$ and backward $I_{c,-}$ critical supercurrents is responsible for SDE.  \label{fig1}}
\end{figure}

The SDE is characterized by the emergence of nonreciprocal charge transport in superconducting systems \cite{RN91,RN218,annurev:/content/journals/10.1146/annurev-conmatphys-032822-033734}. It is explained by the difference of the critical currents $\Delta I_c$ between the forward and backward transport, and the efficiency is defined by $\gamma=(I_{c,+}-|I_{c,-}|)/(I_{c,+}+|I_{c,-}|)$, as shown in Fig. \ref{fig1}. Due to its potential application to the ultra low power quantum rectification devices, SDE has attracted significant attentions both theoretically and experimentally \cite{RN93,RN206,RN80,RN249,RN213,PhysRevLett.128.037001,RN81,RN98,RN208,RN212,RN90,RN89,RN215,RN207,Scammell_2022,RN217,RN211,PhysRevB.106.L140505,PhysRevLett.128.177001,PhysRevB.107.214512,PhysRevLett.131.016001,RN223,PhysRevB.110.024503,PhysRevB.109.064511,RN219,RN220,RN229,RN230,RN227,RN222,RN290,PhysRevB.111.174513}. The SDE can arise in various systems, broadly categorized into Josephson junction configurations \cite{RN93,RN213,RN212} and monolithic superconductors \cite{RN80,PhysRevLett.128.037001}. In the latter systems, this phenomenon is conventionally termed the intrinsic SDE.

The first theoretical studies for the intrinsic SDE was performed in helical superconductors \cite{PhysRevLett.128.037001,RN81,RN98}, in which the spatial-inversion $\mathcal{P}$ symmetry is broken. After breaking the time-reversal $\mathcal{T}$ symmetry by an external magnetic field, they demonstrated SDE can emerge by the cooperation of the Rashba spin orbital coupling (SOC) \cite{EIRashba,PismaZhETF.39.66,PhysRevB.62.4245} and Zeeman term. For most cases, intrinsic SDE is associated with finite momentum superconducting states \cite{PhysRev.135.A550,LO}. Current experimental and theoretical studies demonstrate that simultaneous breaking of $\mathcal{P}$ and $\mathcal{T}$ symmetries constitutes a necessary but insufficient condition for the emergence of the intrinsic SDE \cite{RN218}. Thus, for distinct model systems, numerically intensive calculations are required to determine whether the SDE can be realized. A straightforward universal criterion for diagnosing intrinsic SDE remains lacking. 

In this work, we establish an algebraic diagnostic criterion for the intrinsic SDE. We derive two inequalities that provide a structural condition for nonreciprocity directly from the bare Hamiltonian. A central consequence is that, for systems with $2^N$ degrees of freedom (DOF), these inequalities correspond to a graph-theoretic structure: nonreciprocal models can be systematically constructed from cycles of anticommuting matrix terms, for which the permutation sum is nonzero and given analytically by Bernoulli numbers. This mapping offers a systematic construction scheme for nonreciprocal models, independent of the specific superconducting setup.

The presentation is organized as follows. In Sec.~\ref{sec_foundation} we establish the theoretical foundation for our diagnostic criterion and discuss its regime of validity. In Sec.~\ref{sec_criterion} we introduce the effective time-reversal (ETR) and effective inversion (EI) symmetries, decompose the Hamiltonian into four symmetry sectors, and derive the parity constraints that necessitate simultaneous ETR-EI breaking. We then translate these constraints into two trace inequalities that constitute the central diagnostic criterion. In Sec.~\ref{sec_graph} we map the inequalities to a graph-theoretic problem for $2^N$-DOF systems, evaluate single-cycle contributions analytically in terms of Bernoulli numbers, and develop a systematic generating procedure for nonreciprocal models. In Sec.~\ref{sec_numerical} we provide numerical verification of the graph-theoretic construction. In Sec.~\ref{sec_effective} we extend the criterion to effective bands. We conclude in Sec.~\ref{sec_conclu} with a discussion of the limitations and implications of our results.

\section{Theoretical Foundation and validity \label{sec_foundation}}
The "well-defined bands" denotes models characterized by all bands intersecting the Fermi surface and featuring exclusively effective bilinear terms in the electron field (such as the Rashba model), apart from superconducting interactions. Our analysis shows that the fundamental diagnostic criterion governing the emergence of intrinsic SDE are established through the study of these models.

To describe the SDE phenomenologically, Ginzburg-Landau (GL) theory employs the free energy functional $F\left(\boldsymbol{q}\right)=\alpha_{\boldsymbol{q}}|\Delta_{\boldsymbol{q}}|^2+\left(\beta_{\boldsymbol{q}}/2\right)|\Delta_{\boldsymbol{q}}|^4$, with $\Delta_{\boldsymbol{q}}$ denoting the finite-momentum $\boldsymbol{q}$ pairing order parameter. Non-reciprocity necessitates that $\alpha_{\boldsymbol{q}}$ and $\beta_{\boldsymbol{q}}$ possess contributions odd in $\boldsymbol{q}$ \cite{VictorMEdelstein1996,RN98}, which causes $F(\boldsymbol{q})\neq F(-\boldsymbol{q})$. Our goal is to establish the diagnostic criterion permitting the necessary $\alpha_{\boldsymbol{q}}$ term, because $\alpha_{\boldsymbol{q}}$ constitutes the minimal term for SDE \cite{RN81,RN98}. 

Determined by the specific Hamiltonian model $H$, $\alpha_{\boldsymbol{q}}=1/V-\chi(\boldsymbol{q})$. Here, $V$ is the intensity of superconducting interaction. The $\boldsymbol{q}$-dependent superconducting susceptibility plays a decisive role in it, 
\begin{align}
	\chi(\boldsymbol{q})=k_BT\sum_{\boldsymbol{k},i\omega_n}&\mathrm{Tr}\left[\mathscr{T}^{\dagger}(\hat{\boldsymbol{k}})\mathcal{G}\left(\boldsymbol{k}+\frac{\boldsymbol{q}}{2},i\omega_n\right) \right. \notag \\
	&\left.\times\mathscr{T}(\hat{\boldsymbol{k}})\mathcal{G}^T\left(-\boldsymbol{k}+\frac{\boldsymbol{q}}{2},-i\omega_n\right)\right]. \label{sym}
\end{align}
Within the phenomenological theory of superconductivity \cite{RN77,RN76,RN79}, the form factor for Cooper pairs, denoted by $\mathscr{T}(\hat{\boldsymbol{k}})$, depends on the unit vector $\hat{\boldsymbol{k}}$ in momentum space. The Matsubara Green's function for the electron system described by $H(\boldsymbol{k})$ is given by $\mathcal{G}(\boldsymbol{k},i\omega_n) = \left[i\omega_n - H(\boldsymbol{k})\right]^{-1}$, where $\omega_n = (2n+1)\pi k_B T$ is the fermionic Matsubara frequency at temperature $T$.

\subsection{The asymmetries $\alpha_{\delta\boldsymbol{q}+\boldsymbol{q}_0}\neq\alpha_{-\delta\boldsymbol{q}+\boldsymbol{q}_0}$ and $\alpha_{\boldsymbol{q}}\neq\alpha_{-\boldsymbol{q}}$ \label{sec_alpha}}

As pointed out in Ref.~\cite{PhysRevB.110.024508}, the asymmetry $F(\boldsymbol{q})\neq F(-\boldsymbol{q})$ in the free energy is not sufficient for the emergence of SDE. The zero-current ground state with this free energy asymmetry is a Fulde-Ferrell (FF) state, in which Cooper pairs carry a finite momentum $\boldsymbol{q}_0$. The asymmetry that actually underlies SDE is that of the current-carrying states around this ground state, i.e., $F(\delta\boldsymbol{q}+\boldsymbol{q}_0)\neq F(-\delta\boldsymbol{q}+\boldsymbol{q}_0)$. This distinction reveals an important issue in earlier works \cite{RN81,RN98}: to leading order in the magnetic field $\boldsymbol{B}$, the approximate symmetry $\alpha_{\delta\boldsymbol{q}+\boldsymbol{q}_0}\approx\alpha_{-\delta\boldsymbol{q}+\boldsymbol{q}_0}$ holds, implying that the true SDE arises only at third or higher order in the field \cite{PhysRevLett.128.177001,PhysRevB.110.024508}.

Here we analyze the origin of this approximate symmetry in Rashba systems. To first order in the field $\boldsymbol{B}$, the eigenenergies of the Rashba model are
\begin{equation}
	E_{\boldsymbol{k},\pm}^{\text{R}}\approx\xi_{\boldsymbol{k}}\pm\left[\lambda_Rk+\hat{\boldsymbol{k}}\cdot\left(\boldsymbol{B}\times\hat{\boldsymbol{z}}\right)\right].
\end{equation}
Here $\lambda_R$ is the intensity of the Rashba SOC. The two Fermi surfaces are thus displaced circles, shifted relative to their time-reversal symmetric counterparts. The shifts of the Fermi centers, $\delta\boldsymbol{k}_{\pm}$, are determined by
\begin{align}
	\xi_{{\boldsymbol{k}_{F,\pm}}}\pm\lambda_Rk_{F,\pm}+\boldsymbol{v}_{F,\pm}\cdot \delta\boldsymbol{k}_{\pm}\pm\hat{\boldsymbol{k}}\cdot\left(\boldsymbol{B}\times\hat{\boldsymbol{z}}\right)=0,
\end{align}
where $\boldsymbol{k}_{F,\pm}$ lie on the unperturbed Fermi circles, satisfying $\xi_{{\boldsymbol{k}_{F,\pm}}}\pm\lambda_Rk_{F,\pm}=0$. Using the Fermi velocities
\begin{align}
	\boldsymbol{v}_{F,\pm}=\left.\nabla_{\boldsymbol{k}}\left(\xi_{{\boldsymbol{k}}}\pm\lambda_Rk\right)\right|_{\boldsymbol{k}_{F,\pm}}\equiv v_F\hat{\boldsymbol{k}},
\end{align}
we obtain $\delta\boldsymbol{k}_{\pm}=\mp\left(\boldsymbol{B}\times\hat{\boldsymbol{z}}\right)/v_F$. Crucially, the velocities on the two Fermi circles share the same value $v_F$. As a result, $\delta\boldsymbol{k}_{\pm}$ is independent of the direction $\hat{\boldsymbol{k}}$, meaning that, to linear order in the field, the Fermi circles are rigidly displaced without any distortion.

This rigid displacement is directly responsible for the symmetry $\alpha_{\delta\boldsymbol{q}+\boldsymbol{q}_0}=\alpha_{-\delta\boldsymbol{q}+\boldsymbol{q}_0}$. To illustrate this, we consider a simpler single-band model with a linear shift,
\begin{equation}
	E_{\boldsymbol{k}}^{\text{S}}=\xi_{\boldsymbol{k}}+\lambda k_x. \label{sb}
\end{equation}
The $\boldsymbol{q}$-dependent part of $\alpha_{\boldsymbol{q}}$ is determined by the superconducting susceptibility
\begin{align}
	\chi^{\text{S}}(\boldsymbol{q})&=-k_BT\sum_{\boldsymbol{k},i\omega_n}\frac{1}{i\omega_n-\epsilon-\delta_{\boldsymbol{k}}}\frac{1}{i\omega_n+\epsilon-\delta_{\boldsymbol{k}}} \notag \\
	&=\chi_0+N_F\left\langle\mathcal{C}_0\left(T,\delta_{\boldsymbol{k}}\right)\right\rangle_{\boldsymbol{k}}, \label{q0}
\end{align}
where
\begin{align}
	\epsilon&=\frac{1}{2}\left(E_{\boldsymbol{k}+\boldsymbol{q}/2}^{\text{S}}+E_{-\boldsymbol{k}+\boldsymbol{q}/2}^{\text{S}}\right)\approx\xi_{{\boldsymbol{k}}}+\lambda\frac{q_x}{2}, \\
	\delta_{\boldsymbol{k}}&=\frac{1}{2}\left(E_{\boldsymbol{k}+\boldsymbol{q}/2}^{\text{S}}-E_{-\boldsymbol{k}+\boldsymbol{q}/2}^{\text{S}}\right)\approx v_F\hat{\boldsymbol{k}}\cdot\frac{\boldsymbol{q}}{2}+\lambda k_F\hat{\boldsymbol{k}}\cdot\hat{\boldsymbol{x}}.
\end{align}
Here $\chi_0$ is the superconducting susceptibility for the 2D state, $N_F$ is the density of states (DOS) on the Fermi surface, and $\langle\cdots\rangle_{\boldsymbol{k}}$ denotes the angular average over $\boldsymbol{k}$. The function $\mathcal{C}_0(T,x)=\text{Re}\big[\psi^{(0)}(1/2)-\psi^{(0)}\big(1/2+ix/(2\pi k_BT)\big)\big]$, with $\psi^{(0)}(z)$ the digamma function, can be expanded in even powers of $x$. Setting $\boldsymbol{q}=\delta\boldsymbol{q}+\boldsymbol{q}_0$ with $\boldsymbol{q}_0=-2(\lambda  k_F/v_F)\hat{\boldsymbol{x}}$, Eq.~\eqref{q0} reduces to a function containing only even orders of $\delta\boldsymbol{q}$, yielding $\alpha_{\delta\boldsymbol{q}+\boldsymbol{q}_0}=\alpha_{-\delta\boldsymbol{q}+\boldsymbol{q}_0}$. Clearly, this symmetry originates from the linear form of the $\lambda k_x$ term, which only shifts the Fermi surface without distorting it. If, for instance, this term were replaced by $\lambda k_x^3$, the accidental symmetry would be immediately lifted.

Returning to the Rashba model, the band-dependent splittings become
\begin{align}
	\delta_{\boldsymbol{k}}^{\pm}\approx v_F\hat{\boldsymbol{k}}\cdot\frac{\boldsymbol{q}}{2}\pm\hat{\boldsymbol{k}}\cdot\left(\boldsymbol{B}\times\hat{\boldsymbol{z}}\right).
\end{align}
The $\boldsymbol{q}$-dependent contribution to $\alpha_{\boldsymbol{q}}$ is then
\begin{equation}
	\chi^{\text{R}}(\boldsymbol{q})=\chi_0+N_F^+\left\langle\mathcal{C}_0\left(T,\delta_{\boldsymbol{k}}^+\right)\right\rangle_{\boldsymbol{k}}+N_F^-\left\langle\mathcal{C}_0\left(T,\delta_{\boldsymbol{k}}^-\right)\right\rangle_{\boldsymbol{k}}.
\end{equation}
Setting $\boldsymbol{B}=-B\hat{\boldsymbol{x}}$, the expansion to first order in field reproduces Eq.~(39) of Ref.~\cite{PhysRevB.110.024508}. The resulting approximate symmetry $\alpha_{\delta\boldsymbol{q}+\boldsymbol{q}_0}\approx\alpha_{-\delta\boldsymbol{q}+\boldsymbol{q}_0}$ relies on the integral identity
\begin{equation}
	\int_0^{2\pi}\cos^{2n-1}(\theta-\theta')\sin\theta\,d\theta=\sin\theta'\int_0^{2\pi}\cos^{2n}(\theta-\theta')\,d\theta,
\end{equation}
which holds precisely because of the linear-$\boldsymbol{k}$-dependence of the Rashba term.

Therefore, the inconsistency between the two asymmetries $\alpha_{\delta\boldsymbol{q}+\boldsymbol{q}_0}\neq\alpha_{-\delta\boldsymbol{q}+\boldsymbol{q}_0}$ and $\alpha_{\boldsymbol{q}}\neq\alpha_{-\boldsymbol{q}}$ is attributed to the rigid band shift that occurs in simple, low-order models. In real materials, however, symmetry-breaking terms are not restricted to the lowest order, as in $\boldsymbol{k}\cdot\boldsymbol{p}$ approximations. The generic presence of higher-order $\boldsymbol{k}$-dependent symmetry-breaking terms ensures that, in most physically relevant cases, the two asymmetries are consistent.

\begin{widetext}
	We now extend this analysis to multi-band systems, which reveals even stronger constraints on the conditions under which the two asymmetries can be inconsistent. Consider a set of bands labeled by index $s$, with eigenenergies of the form
	\begin{equation}
		E_{\boldsymbol{k},s}^{\text{M}}=\xi_{\boldsymbol{k},s}+\lambda_s k_x,
	\end{equation}
	where $\lambda_s$ is a band-dependent factor that quantifies the strength of the symmetry-breaking term in each band. The total $\alpha_{\boldsymbol{q}}$ is obtained by summing over all bands:
	\begin{align}
		\alpha_{\boldsymbol{q}}&=\sum_{s,n}N_F^sA_n\left\langle\left(v_F^s\hat{\boldsymbol{k}}\cdot\boldsymbol{q}+2\lambda_s\boldsymbol{k}\cdot\hat{\boldsymbol{x}}\right)^{2n}\right\rangle_{\boldsymbol{k}} \notag \\
		&\approx\sum_{s,n}\int_0^{2\pi}d\theta_{\boldsymbol{k}}\,N_F^sA_n\left[v_F^sq\cos\left(\theta_{\boldsymbol{k}}-\theta_{\boldsymbol{q}}\right)\right]^{2n-1}\left[v_F^sq\cos\left(\theta_{\boldsymbol{k}}-\theta_{\boldsymbol{q}}\right)+4n\lambda_sk_F\cos\theta_{\boldsymbol{k}}\right]. \label{multi_alpha}
	\end{align}
	Here we have expanded to first order in $\lambda_s$. Following the approach of Ref.~\cite{PhysRevB.110.024508}, we examine the asymmetry by evaluating $\alpha_{\pm\delta\boldsymbol{q}+\boldsymbol{q}_0}$:
	\begin{align}
		\alpha_{\pm\delta\boldsymbol{q}+\boldsymbol{q}_0}\approx\sum_{s,n}\int_0^{2\pi}d\theta_{\boldsymbol{k}}\,N_F^sA_n\left[v_F^s\delta q\cos\left(\theta_{\boldsymbol{k}}-\theta_{\boldsymbol{q}}\right)\right]^{2n-1}\left[v_F^s\left(\delta q\pm2nq_0\right)\cos\left(\theta_{\boldsymbol{k}}-\theta_{\boldsymbol{q}}\right)\pm4n\lambda_sk_F\cos\theta_{\boldsymbol{k}}\right],
	\end{align}
	where $q_0$ is treated as $\mathcal{O}(\lambda_s)$. The terms odd in $\delta\boldsymbol{q}$, which are responsible for the SDE, are:
	\begin{align}
		&\sum_{s,n}\int_0^{2\pi}d\theta_{\boldsymbol{k}}\,N_F^sA_n\left[v_F^s\delta q\cos\left(\theta_{\boldsymbol{k}}-\theta_{\boldsymbol{q}}\right)\right]^{2n-1}\left[2nq_0v_F^s\cos\left(\theta_{\boldsymbol{k}}-\theta_{\boldsymbol{q}}\right)+4n\lambda_sk_F\cos\theta_{\boldsymbol{k}}\right] \notag \\
		&=\sum_{n}A_n\,\delta q^{2n-1}\,2n\left[\int_{0}^{2\pi}d\theta_{\boldsymbol{k}}\cos^{2n}\left(\theta_{\boldsymbol{k}}-\theta_{\boldsymbol{q}}\right)\right]\sum_{s}N_F^s (v_F^s)^{2n-1}\left(q_0 v_F^s+2\lambda_sk_F\cos\theta_{\boldsymbol{q}}\right).
	\end{align}
\end{widetext}

The SDE is absent, and the system supports a pure FF state without nonreciprocal critical currents, only when this odd contribution vanishes identically for every integer $n$. This requires
\begin{equation}
	\sum_{s}N_F^s (v_F^s)^{2n-1}\left(q_0 v_F^s+2\lambda_sk_F\cos\theta_{\boldsymbol{q}}\right)=0,
\end{equation}
which can be recast as
\begin{align}
	\frac{\sum_{s}N_F^s(\lambda_s/v_F^s)\left(v_F^s\right)^{2n}}{\sum_{s}N_F^s\left(v_F^s\right)^{2n}}=\text{Const},
\end{align}
where the constant must be independent of $n$. This condition is satisfied if either of the following holds: (i) the ratio $\lambda_s/v_F^s$ is identical across all bands; (ii) the Fermi velocity $v_F^s=v_F$ is band-independent.

Condition (i) is the multi-band analogue of the single-band case analyzed earlier: all bands undergo precisely the same rigid shift, which forces the odd terms in $\delta\boldsymbol{q}$ to cancel exactly, leading to a strict absence of SDE. Condition (ii) yields an approximate equivalence: with a common Fermi velocity, the odd terms cancel at order $\mathcal{O}(\lambda_s)$, but this cancellation is not exact beyond the perturbative expansion. The Rashba model satisfies condition (ii), which explains the approximate symmetry $\alpha_{\delta\boldsymbol{q}+\boldsymbol{q}_0}\approx\alpha_{-\delta\boldsymbol{q}+\boldsymbol{q}_0}$ found in it.

In generic multi-band materials, however, neither condition is expected to hold. The ratio $\lambda_s/v_F^s$ varies from band to band, and Fermi velocities differ across orbital characters. Consequently, the cancellation of the odd terms is nongeneric, and the SDE is expected to emerge whenever $\alpha_{\boldsymbol{q}}\neq\alpha_{-\boldsymbol{q}}$. In other words, the two asymmetries $F(\boldsymbol{q})\neq F(-\boldsymbol{q})$ and $F(\delta\boldsymbol{q}+\boldsymbol{q}_0)\neq F(-\delta\boldsymbol{q}+\boldsymbol{q}_0)$ are qualitatively equivalent in all physically realistic situations. The condition $\alpha_{\boldsymbol{q}}\neq\alpha_{-\boldsymbol{q}}$, which arises from terms odd in $\boldsymbol{q}$, is therefore sufficient to establish the existence of the SDE in real systems. This justifies the use of this simpler condition, rather than the more refined one $\alpha_{\delta\boldsymbol{q}+\boldsymbol{q}_0}\neq\alpha_{-\delta\boldsymbol{q}+\boldsymbol{q}_0}$, as the foundation for the algebraic diagnostic criterion developed in this work.

\subsection{Theoretical-limit $\alpha_{\boldsymbol{q}}$}
\begin{figure}[htp]
	\centering
	\includegraphics[width=1.0\columnwidth]{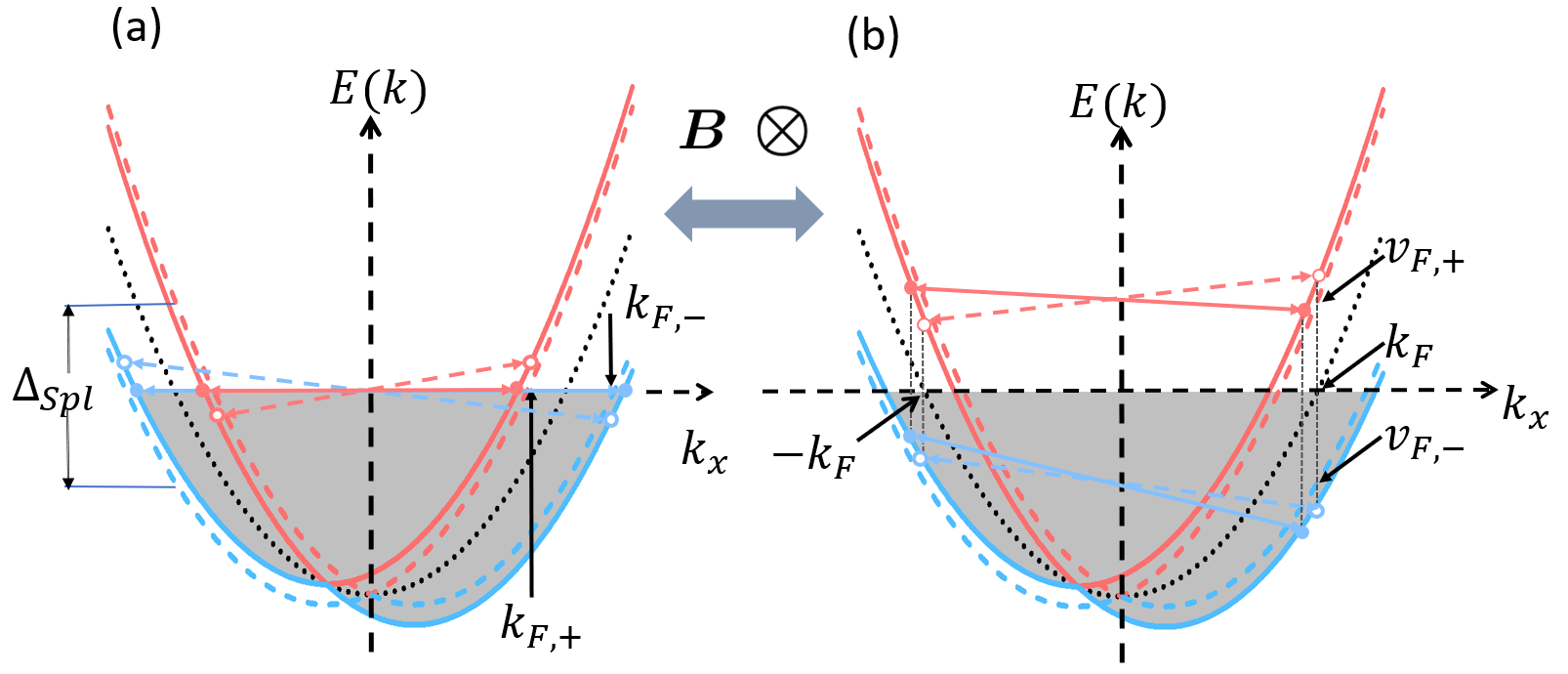}
	\caption{The microscopic origin of FF state in Rashba system, where the x-axis is along the current. $\Delta_{Spl}$ measures the band spliting from Rashba SOC. (a) physical-limit ($\Delta_{\boldsymbol{q}}\ll\Delta_{Spl}$) and (b) theoretical-limit ($\Delta_{\boldsymbol{q}}\gg\Delta_{Spl}$). The red curvature is the inner band, while the blue one is the outer band. The black dotted, red and blue dashed, and red and blue solid curvatures correspond to the dispersion without $\mathcal{T}$ and $\mathcal{P}$, with only  $\mathcal{P}$, and with both $\mathcal{T}$ and $\mathcal{P}$ symmetry breakings, respectively. The red and blue solid and dashed lines with arrows connecting two $\circ$ and two $\bullet$ denote the construction of Cooper pairs. $k_{F,s}$ denotes the Fermi momentum for band indexed by $s$ and $v_{F,s}$ indicates the relevant velocity.  Note that all $k_{F,s}$ and $v_{F,s}$ correspond solely to the case with only $\mathcal{P}$ symmetry breaking. \label{fig_aq}}
\end{figure}

The conventional approach to evaluating Eq. (\ref{sym}) adopts the physical limit where the superconducting pairing strength $\Delta_{\boldsymbol{q}}$ is much smaller than the band splitting $\Delta_{Spl}$ [i.e., $\Delta_{\boldsymbol{q}} \ll \Delta_{Spl}$, Fig. \ref{fig_aq}(a)]. In this regime, bands distant from the Fermi surface contribute negligibly to superconductivity, resulting in an incomplete trace in Eq. (\ref{sym}) (see Appendix \ref{standard}). In contrast, we postulate the opposite, theoretical limit ($\Delta_{\boldsymbol{q}} \gg \Delta_{Spl}$, Fig. \ref{fig_aq}(b)), where the split bands are approximated by a single Fermi surface. This enables a complete trace evaluation of Eq. (\ref{sym}) (Appendix \ref{fake}), a crucial step for deriving the algebraic diagnostic criterion. Crucially, for identifying nonreciprocity, these two limits are qualitatively equivalent. This equivalence stems from the fact that the nonreciprocal response originates from the intrinsic asymmetry of the electronic bands, not from the magnitude of the pairing strength.

To illustrate, we use the emergence of FF state for example. In the band basis, the nonreciprocity-contributing term in $\alpha_{\boldsymbol{q}}$ to the linear-$\boldsymbol{q}$ order can be expressed as a weighted sum over all bands $[N_{F,s}v_{F,s}(k_{F,s}) \Delta E_s]$ (Appendix \ref{simplified}), where $N_{F,s}$ is the band-dependent DOS, $v_{F,s}(k_{F,s})$ is the Fermi velocity at Fermi wave vector $k_{F,s}$, and $\Delta E_s$ is the energy difference between 
$\pm \boldsymbol{k}$ on band $s$. Using the Rashba model as an example [with $\Delta_{\boldsymbol{q}}\ll \Delta_{Spl} $, Fig.  \ref{fig_aq} (a)], the FF state originates from the band antisymmetry about zero momentum.  In Fig.  \ref{fig_aq} (a), the net nonreciprocity arises from the imbalance in $N_{F,s}$ between two split bands \cite{RN81}. Since $v_{F,+}(k_{F,+})\approx v_{F,-}(k_{F,-})$, $\Delta E_{+}\approx-\Delta E_{-}$ and  $N_{F,s} \propto k_{F,s}/v_{F,s}(k_{F,s})$,
the dominant linear-$\boldsymbol{q}$-dependent contribution to $\alpha_{\boldsymbol{q}}$ is proportional to $\sum_s k_{F,s}\Delta E_s$ and is governed by the lower band with the larger $k_F$ [Fig. \ref{fig_aq}(a)].

Extending analysis to the opposite limit $\Delta_{\boldsymbol{q}}\gg \Delta_{Spl} $ limit [Fig.  \ref{fig_aq} (b)], the approximation of a single Fermi surface [dashed parabolic band in Fig. \ref{fig_aq} (b)] simplifies the linear-$\boldsymbol{q}$-dependent term in $\alpha_{\boldsymbol{q}}$ (referred to here as the theoretical-limit $\alpha _{\boldsymbol{q}}$), making it proportional to $\sum_s v_{F,s}(k_{F})\Delta E_s$. Then, the higher band with the larger $v_F$ dominates [Fig. \ref{fig1}(d)]. This behavior arises because the minimization of free energy drives the system toward FF state, where the pairing momentum $\boldsymbol{q}$ takes non-zero value to compensate for the energy differences between electrons at  $\pm \boldsymbol{k}$ [manifested as the evolution from open-circle to solid-circle pairings in Fig. \ref{fig_aq} (b)].

Despite differences in the specific expressions where physical-limit $\alpha_{\boldsymbol{q}}$ [Fig. \ref{fig_aq}(a)] and theoretical-limit $\alpha _{\boldsymbol{q}}$ [Fig. \ref{fig_aq}(b)] may differ in magnitude and even sign (Appendices \ref{standard} and \ref{fake}), both regimes exhibit nonreciprocity. Therefore, the theoretical-limit $\alpha _{\boldsymbol{q}}$ provides a qualitatively valid foundation for deriving a general diagnostic criterion for the emergence of the SDE when higher-order $\boldsymbol{q}$-terms are incorporated. 

\section{Criterion of SDE for well-defined bands \label{sec_criterion}}
\subsection{Effective symmetries and Hamiltonian decomposition}
Equipped with the previous foundations, the diagnostic criterion for SDE is established starting from Eq.~(\ref{sym}) in the theoretical-limit. We first introduce two effective symmetries, motivated by the following considerations. In most studies of $s$-wave spin-singlet pairing, the pairing form factor is $\mathscr{T}=i\sigma^y$. Commuting this factor to the right of the trace in the superconducting susceptibility reveals an apparent connection between the SDE and the physical time-reversal symmetry, conventionally defined as $\mathcal{T}=i\sigma^y\mathcal{K}$, where $\mathcal{K}$ denotes complex conjugation. This identification, however, is not universally applicable: when the system possesses additional DOF---such as orbitals, sublattices, or valleys---or when the pairing is unconventional, both the physical time-reversal operator $\mathcal{T}$ and the pairing form factor $\mathscr{T}$ must be suitably generalized, and the two no longer coincide.

Nevertheless, the symmetry properties of the superconducting susceptibility are fundamentally tied to the pairing form factor $\mathscr{T}$, not to the physical time-reversal symmetry. This connection persists regardless of the number of DOF. We therefore introduce an "effective time-reversal" (ETR) transformation directly associated with $\mathscr{T}$. This transformation, denoted by $\hat{T} \equiv \mathscr{T}\mathcal{K}$, combines the momentum reversal $\bm{k} \to -\bm{k}$ with the matrix part inherited from the pairing. To generalize the analysis, we consider an arbitrary pairing form factor $\mathscr{T}(\hat{\boldsymbol{k}})=i\Phi(\hat{\boldsymbol{k}})\sigma^y$. Our framework applies whenever $\hat{T}=\mathscr{T}(\hat{\boldsymbol{k}})\mathcal{K}$ defines a well-defined transformation on the model's Hilbert space with eigenvalues $\pm 1$, in direct analogy to the conventional time-reversal operation. 

\begin{widetext}
	The Hamiltonian then decomposes into ETR-symmetric and ETR-antisymmetric components via $\hat{T}^\dagger H(-\boldsymbol{k}) \hat{T} = H^{TS}(\boldsymbol{k}) - H^{TA}(\boldsymbol{k})$, from which we obtain
	\begin{align}
		\mathscr{T}\mathcal{G}^T\left(-\boldsymbol{k}+\frac{\boldsymbol{q}}{2},-i\omega_n\right)\mathscr{T}^{\dagger}=-\left[i\omega_n+H^{TS}\left(\boldsymbol{k}-\frac{\boldsymbol{q}}{2}\right)-H^{TA}\left(\boldsymbol{k}-\frac{\boldsymbol{q}}{2}\right)\right]^{-1}.
	\end{align}
	This implies the $\boldsymbol{q}$-expansion of $\alpha_{\boldsymbol{q}}$ further depends on the symmetry $\bm{k}\to-\bm{k}$, which is not necessarily identical to the physical spatial inversion $\mathcal{P}$. While $\mathcal{P}$ reverses momentum, it may also act nontrivially on internal DOF; for instance, interchanging the two valleys in a valley system. To isolate the parity properties of the Hamiltonian under momentum reversal, we define the "effective inversion" (EI) operation $\hat{I}: \bm{k}\to-\bm{k}$, which acts purely on momentum. All operations on the spinor DOF are thereby absorbed into the ETR transformation. We assume a general power-series expansion for the Hamiltonian, $H(\boldsymbol{k}) = \sum_{n_x,n_y} k_x^{n_x} k_y^{n_y} \mathcal{A}_{n_x,n_y}$, where $n_x, n_y$ are nonnegative integers and $\mathcal{A}$ denotes a $\boldsymbol{k}$-independent matrix whose dimension matches that of the effective spinor space. The $\boldsymbol{q}$-dependence is then expressed as
	\begin{align}
		H\left(\boldsymbol{k}+\frac{\boldsymbol{q}}{2}\right)&=\sum_{l_x,l_y,n_x,n_y}\left(\frac{1}{2}\right)^{l_x+l_y}\binom{n_x+l_x}{l_x}\binom{n_y+l_y}{l_y}q_x^{l_x}q_y^{l_y}k_x^{n_x}k_y^{n_y}\mathcal{A}_{l_x+n_x,l_y+n_y} \notag \\
		&\equiv\sum_{l_x,l_y,n_x,n_y}q_x^{l_x}q_y^{l_y}k_x^{n_x}k_y^{n_y}\mathcal{A}_{l_x,n_x;l_y,n_y}.
	\end{align}
	
	With respect to ETR and EI symmetries, the Hamiltonian decomposes into four distinct sectors: $H^{TS,IS}$, $H^{TS,IA}$, $H^{TA,IS}$, and $H^{TA,IA}$, where the superscripts $S$ ($A$) denote symmetry (antisymmetry) under the corresponding operation. For well-defined models, one can always isolate a nonzero term from $H^{TS,IS}$ that is proportional to the identity matrix $\boldsymbol{1}$; we denote this term by $E^{TS,IS}$, which yields a single symmetric Fermi surface. The product inside the trace of Eq. (\ref{sym}) can then be cast as:
	\begin{equation}
		\left[i\omega_n-E^{TS,IS}\left(\boldsymbol{k}\right)-\left(\delta_1+\delta_2\right)\right]^{-1}\left[i\omega_n+E^{TS,IS}\left(\boldsymbol{k}\right)-\left(\delta_1-\delta_2\right)\right]^{-1}, \label{summation}
	\end{equation}
	where
	\begin{equation}
		E^{TS,IS} \subset H^{TS,IS}, \quad H^{TS,IS} = \sum_{n_x+n_y=2m} k_x^{n_x} k_y^{n_y} \mathcal{A}^{TS}, \quad E^{TS,IS} \propto \boldsymbol{1}.
	\end{equation}
	The remaining terms are absorbed into the quantities 
	\begin{align}
		\delta_{1}&=\sum_{\substack{n_x,n_y\\l_x+ly=2m+1}}q_x^{l_x}q_y^{l_y}k_x^{n_x}k_y^{n_y}\mathcal{A}^{TS}+\sum_{\substack{n_x,n_y\\l_x+ly=2m}}q_x^{l_x}q_y^{l_y}k_x^{n_x}k_y^{n_y}\mathcal{A}^{TA}, \\
		\delta_{2}&=\sum_{\substack{n_x,n_y\\l_x+ly=2m}}q_x^{l_x}q_y^{l_y}k_x^{n_x}k_y^{n_y}\mathcal{A}^{TS}+\sum_{\substack{n_x,n_y\\l_x+ly=2m+1}}q_x^{l_x}q_y^{l_y}k_x^{n_x}k_y^{n_y}\mathcal{A}^{TA}.
	\end{align}
	For brevity, we suppress the subscripts $l_x,l_y,n_x,n_y$ on the tensor matrices $\mathcal{A}$, as only the qualitative structure matters for our purpose. Here $m$ is a nonnegative integer, and the first term in $\delta_2$ should be understood to exclude those contributions already included in $E^{TS,IS}$ when $m=0$. 
\end{widetext}

The analysis proceeds in a similar fashion for valleytronic Hamiltonians. Such a Hamiltonian splits into valley-symmetric and valley-antisymmetric components,
\begin{align}
	H_{\text{valley}}(\boldsymbol{k}) = H_o(\boldsymbol{k}) + \eta^z H_v(\boldsymbol{k}),
\end{align}
where $H_o(\boldsymbol{k})$ collects the valley-symmetric terms, $H_v(\boldsymbol{k})$ contains the valley-antisymmetric contributions, and $\eta^z = \pm 1$ is the valley Pauli matrix labeling states near the $\pm\boldsymbol{K}$ valleys. Cooper pairs in these systems are formed by electrons from opposite valleys; for instance, the singlet configuration reads $c_{-\boldsymbol{K}-\boldsymbol{k},\downarrow}c_{\boldsymbol{K}+\boldsymbol{k},\uparrow}-c_{-\boldsymbol{K}-\boldsymbol{k},\uparrow}c_{\boldsymbol{K}+\boldsymbol{k},\downarrow}-c_{\boldsymbol{K}-\boldsymbol{k},\uparrow}c_{-\boldsymbol{K}+\boldsymbol{k},\downarrow}+c_{\boldsymbol{K}-\boldsymbol{k},\downarrow}c_{-\boldsymbol{K}+\boldsymbol{k},\uparrow}$. Defining the electron field as $c_{\boldsymbol{k},\eta,\sigma}\equiv c_{\eta\boldsymbol{K}+\boldsymbol{k},\sigma}$, this pairing takes the form $i\eta^x\sigma^y$, where $\sigma^y$ acts in spin space. Equation~(\ref{sym}) can still be employed to determine the symmetry conditions for SDE, provided the pairing form factor is replaced by $\mathscr{T}(\hat{\boldsymbol{k}})=i\Phi(\hat{\boldsymbol{k}})\eta^x\sigma^y$. The Green's function becomes $\mathcal{G}(\boldsymbol{k}, i\omega_n) = [i\omega_n - H_{\text{valley}}(\boldsymbol{k})]^{-1}$. It is important to note that the label "$IS$" in the valleytronic context does not refer to physical inversion $\mathcal{P}$ (which acts as $\mathcal{P} H(\boldsymbol{k}) \mathcal{P} = \eta^x H(-\boldsymbol{k}) \eta^x$), but rather to the effective inversion (EI) operation $\hat{I}$ defined by $\hat{I} H(\boldsymbol{k}) \hat{I} = H(-\boldsymbol{k})$. We adopt this convention so that the SDE criterion retains a unified form for both generic and valleytronic cases. With this unified convention, the valleytronic case reduces to the same algebraic structure as the generic multi-band case analyzed above. The symmetry analysis that follows therefore applies to both settings on equal footing.

\subsection{Symmetry constraints and the necessity of simultaneous ETR-EI breaking}
Introducing the Green's functions $\mathcal{G}_{e(h)} = [i\omega_n \mp E^{TS,IS}]^{-1}$, which are likewise proportional to $\boldsymbol{1}$, the full product Eq. (\ref{summation}) becomes $\mathcal{G}_h [1 - \mathcal{G}_e(\delta_1+\delta_2) - (\delta_1-\delta_2)\mathcal{G}_h + \mathcal{G}_e(\delta_1+\delta_2)(\delta_1-\delta_2)\mathcal{G}_h]^{-1} \mathcal{G}_e$. Owing to the crucial identity Eq. (\ref{Chilm}), any nonvanishing contribution must be expanded in even powers of $\mathcal{G}$. Consequently, the problem reduces to determining whether $(\delta_1+\delta_2)^{n_+} (\delta_1-\delta_2)^{n_-}$ yields an odd-order term in $\boldsymbol{q}$, where $n_+$ and $n_-$ are integers satisfying $n_+ + n_- = \text{even}$. The existence of such terms signals the emergence of nonreciprocity and hence the presence of intrinsic SDE.

To make the symmetry constraints explicit, we decompose the ETR operation into two steps: $\hat{T}^\dagger H(-\boldsymbol{k}) \hat{T} = \hat{T}^\dagger \hat{I} H(\boldsymbol{k}) \hat{I} \hat{T}$. Here $\hat{I}$ handles the momentum reversal, so that the remaining $\hat{T}^\dagger (\cdots) \hat{T}$ acts on $H(\boldsymbol{k})$ without changing the sign of $\boldsymbol{k}$. The four Hamiltonian sectors transform under this operation as
\begin{equation}
	\hat{T}^\dagger\mathcal{A}^{TS,IS}\hat{T}=\mathcal{A}^{TS,IS},\quad\hat{T}^\dagger\mathcal{A}^{TA,IA} \hat{T}=\mathcal{A}^{TA,IA}, \label{prop1}
\end{equation}
\begin{equation}
	\hat{T}^\dagger\mathcal{A}^{TS,IA}\hat{T}=-\mathcal{A}^{TS,IA},\quad\hat{T}^\dagger\mathcal{A}^{TA,IS} \hat{T}=-\mathcal{A}^{TA,IS}. \label{prop2}
\end{equation}
The remaining operation is decomposed as $\hat{T} = \mathscr{T}\mathcal{K}$, where $\mathscr{T}$ is the unitary matrix part and $\mathcal{K}$ denotes complex conjugation. 

For nondissipative systems, $\alpha_{\boldsymbol{q}}$ is real, and is therefore invariant under $\mathcal{K}$. This invariance guarantees that the trace expression is unchanged under the operations that follow. The trace $\mathrm{Tr}[(\delta_1+\delta_2)^{n_+} (\delta_1-\delta_2)^{n_-}]$ contains an even number of matrix factors. Because $\mathscr{T}$ is unitary, $\mathscr{T}\mathscr{T}^{\dagger} = \boldsymbol{1}$ holds exactly. Inserting this identity between adjacent factors and rebracketing yields an expression in which each matrix factor $M_i$ is conjugated as $\mathscr{T}^{\dagger} M_i \mathscr{T}$. Together with the action of $\mathcal{K}$, this is precisely the ETR transformation $\hat{T}$ applied to the individual factor. Products containing an odd number of factors that are odd under $\hat{T}$ therefore acquire an overall minus sign and must vanish. The sum over all permutations of the matrix factors, which is an essential ingredient of the graph-theoretic framework developed later, absorbs the reversal of factor ordering introduced by $\mathcal{K}$, making the above argument fully consistent.

This parity constraint leads directly to the conclusion that breaking either ETR or EI symmetry alone is insufficient to generate nonreciprocity. We prove this by examining each case.

First, consider systems that preserve ETR symmetry but break EI. In this case $H^{TA}=0$, so the Hamiltonian contains only the $H^{TS,IS}$ and $H^{TS,IA}$ sectors. All terms containing $\mathcal{A}^{TA}$ disappear from the definitions of $\delta_1$ and $\delta_2$. By construction, $\delta_1$ then contains only odd powers of $\boldsymbol{q}$, while $\delta_2$ contains only even powers of $\boldsymbol{q}$. Any term contributing to $\alpha_{\boldsymbol{q}}$ is a product of the form $(\delta_1+\delta_2)^{n_+}(\delta_1-\delta_2)^{n_-}$. For each term in the expansion, after factoring out even powers of each $\delta_\nu$ ($\nu\in\{1,2\}$), a nonvanishing odd-order contribution in $\boldsymbol{q}$ necessarily requires a factor of $\delta_1\delta_2$. For this product to survive the Fermi-surface average, the total power of $\boldsymbol{k}$ must be even, which demands that $\delta_1$ ($\delta_2$) and $\delta_2$ ($\delta_1$) originate from $H^{TS,IA}$ and $H^{TS,IS}$, respectively. But this introduces an odd number of $H^{TS,IA}$ factors, violating the parity constraint of the $\hat{T}$ transformation. Hence, in ETR-preserving systems, $\alpha_{\boldsymbol{q}}$ contains only even powers of $\boldsymbol{q}$, and no SDE can arise.

Next, consider systems that preserve EI symmetry but break ETR. Under EI symmetry, $H^{IA}=0$, so the Hamiltonian contains only terms that are even in $\boldsymbol{k}$. When $H(\boldsymbol{k}+\boldsymbol{q}/2)$ is expanded, the powers of $\boldsymbol{q}$ and $\boldsymbol{k}$ in each term share the same parity. The product structure of the superconducting susceptibility then forces any nonzero contribution to have $\boldsymbol{q}$ and $\boldsymbol{k}$ both even or both odd. Nonreciprocity requires an odd total power of $\boldsymbol{q}$, which would force an odd total power of $\boldsymbol{k}$ as well. However, the Fermi-surface average $\langle\cdots\rangle_{\boldsymbol{k}}$ is nonzero only for even total powers of $\boldsymbol{k}$. The two requirements are incompatible. Thus, in EI-preserving systems, $\alpha_{\boldsymbol{q}}$ again contains only even powers of $\boldsymbol{q}$.

In summary, SDE can emerge only when both ETR and EI symmetries are simultaneously broken; specifically, through the coexistence of $H^{TS,IA}$ and $H^{TA,IS}$ (first channel), or $H^{TS,IS}$ and $H^{TA,IA}$ (second channel) in the expansion. This is the algebraic origin of the well-known requirement that both time-reversal and inversion symmetries must be broken for intrinsic SDE. We emphasize, however, that simultaneous breaking is necessary but not sufficient; the precise nonreciprocity conditions are determined by the trace inequalities, Eqs.~(\ref{Eq1}) and~(\ref{Eq2}).

\begin{table*}[ptb]
	\renewcommand{\arraystretch}{1.5}
	\caption{Classification of terms according to the parities of $\boldsymbol{q}$ ($O_{\boldsymbol{q}}$) and $\boldsymbol{k}$ ($O_{\boldsymbol{k}}$). We use $o$ and $e$ to denote odd and even orders, respectively. The second and third columns list the terms appearing at second order in $\delta$.}
	\label{Ts}
	\begin{ruledtabular}
		\begin{tabular}{ccc}
			$\{O_{\boldsymbol{q}}, O_{\boldsymbol{k}}\}$ & $\delta_\nu^2 \; (\nu \in \{1,2\})$ & $\delta_1\delta_2$ \\
			\hline
			$\{o,e\}$ & $\mathcal{A}^{TS,IS}\mathcal{A}^{TA,IA}$, $\mathcal{A}^{TS,IA}\mathcal{A}^{TA,IS}$ & $\mathcal{A}^{TS,IS}\mathcal{A}^{TS,IA}$, $\mathcal{A}^{TA,IA}\mathcal{A}^{TA,IS}$ \\
			\hline
			$\{e,o\}$ & $\mathcal{A}^{TS,IS}\mathcal{A}^{TS,IA}$, $\mathcal{A}^{TA,IA}\mathcal{A}^{TA,IS}$ & $\mathcal{A}^{TS,IS}\mathcal{A}^{TA,IA}$, $\mathcal{A}^{TS,IA}\mathcal{A}^{TA,IS}$ \\
			\hline
			\multirow{2}*{$\{o,o\}$} & \multirow{2}*{$\mathcal{A}^{TS,IS}\mathcal{A}^{TA,IS}$, $\mathcal{A}^{TS,IA}\mathcal{A}^{TA,IA}$} & $\mathcal{A}^{TS,IS}\mathcal{A}^{TS,IS}$, $\mathcal{A}^{TS,IA}\mathcal{A}^{TS,IA}$, \\
			& & $\mathcal{A}^{TA,IS}\mathcal{A}^{TA,IS}$, $\mathcal{A}^{TA,IA}\mathcal{A}^{TA,IA}$ \\
			\hline
			\multirow{2}*{$\{e,e\}$} & $\mathcal{A}^{TS,IS}\mathcal{A}^{TS,IS}$, $\mathcal{A}^{TS,IA}\mathcal{A}^{TS,IA}$, & \multirow{2}*{$\mathcal{A}^{TS,IS}\mathcal{A}^{TA,IS}$, $\mathcal{A}^{TS,IA}\mathcal{A}^{TA,IA}$} \\
			& $\mathcal{A}^{TA,IS}\mathcal{A}^{TA,IS}$, $\mathcal{A}^{TA,IA}\mathcal{A}^{TA,IA}$ &
		\end{tabular}
	\end{ruledtabular}
\end{table*}

\subsection{Two inequalities that govern the emergence of intrinsic SDE}
The parity constraints derived in the previous subsection dictate which combinations of Hamiltonian sectors can yield a nonvanishing trace. Table~\ref{Ts} summarizes these contributions at second order in $\delta_{\nu}$, $\nu\in\{1,2\}$. The entries are exhaustive for arbitrary $(n_+, n_-)$, because higher-order contributions are obtained by multiplying the second-order building blocks by factors of the form $\prod_{\zeta,\zeta' \in \{S,A\}} (\mathcal{A}^{T\zeta,I\zeta'})^{2m_{\zeta\zeta'}}$, which are even in both $\boldsymbol{q}$ and $\boldsymbol{k}$ and therefore do not alter the parity classification.

Table~\ref{Ts} reveals a systematic pattern. The SDE requires an odd total power of $\boldsymbol{q}$ and an even total power of $\boldsymbol{k}$. Consulting the table, only the $\{o,e\}$ row satisfies this condition. Reading off the corresponding matrix structures, the model must satisfy at least one of the following inequalities:
\begin{widetext}
	\begin{align}
		\Pi_1\left(P\right)\equiv \mathrm{Tr}\left\{P\left[\prod_{\zeta,\zeta'\in\{S,A\}}\left(H^{T\zeta,I\zeta'}\right)^{2n_{\zeta\zeta'}}H^{TS,IA}H^{TA,IS}\right]\right\}\neq 0, \label{r1} \\
		\Pi_2\left(P\right)\equiv \mathrm{Tr}\left\{P\left[\prod_{\zeta,\zeta'\in\{S,A\}}\left(H^{T\zeta,I\zeta'}\right)^{2n_{\zeta\zeta'}}H^{TS,IS}H^{TA,IA}\right]\right\}\neq 0. \label{r2}
	\end{align}
\end{widetext}
Here, the integer $n_{\zeta\zeta'}$s are definite numbers for the given model, which produces enough forms of the matrices \cite{note}. For most simple models~\cite{PhysRevLett.128.037001,RN81,RN98,PhysRevB.106.L140505,PhysRevLett.133.216201,Scammell_2022,RN219,PhysRevB.110.024503,PhysRevB.109.064511}, setting $n_{\zeta\zeta'}=0$ suffices. $P$ denotes the permutation of the terms. Although most existing models are relatively simple (e.g., the models listed in Table~\ref{table}), the Hamiltonian terms need not commute with one another. Consequently, different permutations can, in principle, yield distinct contributions. Therefore, Eqs.~(\ref{r1}) and (\ref{r2}) should be regarded as constraints beyond the basic symmetry requirements, rather than as the definitive criterion for the emergence of intrinsic SDE. To ensure a finite net SDE efficiency, one must verify that the contributions from distinct permutations do not cancel exactly.

\begin{table*}[ptb]
	\renewcommand{\arraystretch}{1.5}
	\caption{The terms for different models are categorized in the sense of our ETR and EI symmetry definitions. The second column is for the form factor of the considered Cooper pair. Columns third to sixth are the terms' classifications with the decided integer $n_{\zeta\zeta'}$. The functions in the fifth row are $\xi_{\boldsymbol{k}}^e=-\sum_{j=1}^3t_j\cos(\boldsymbol{a}_j\cdot\boldsymbol{k})$, $\xi_{\boldsymbol{k}}^o=-\sum_{j=1}^3t_j\sin(\boldsymbol{a}_j\cdot\boldsymbol{k})$, $A_{\pm}=(\cos\phi_+\pm\cos\phi_-)/2$, $B_{\pm}=(\sin\phi_+\pm\sin\phi_-)/2$, $\mu=(\mu_++\mu_-)/2$, $\delta\mu=(\mu_+-\mu_-)/2$. }
	\label{table}
	\begin{ruledtabular}
		\begin{tabular}{cccccc}
			
			Model  &  ETR($\Phi$)  &  $H^{TS,IS}\left[n_{SS}\right]$ & $H^{TS,IA}\left[n_{SA}\right]$  & $H^{TA,IS}\left[n_{AS}\right]$ & $H^{TA,IA}\left[n_{AA}\right]$ \tabularnewline
			\hline 
			Rashba\cite{PhysRevLett.128.037001,RN81,RN98} & $\boldsymbol{1}$ & $\xi_{\boldsymbol{k}}\;[0]$ & $\lambda_R\left(\boldsymbol{k}\times\boldsymbol{\sigma}\right)\cdot\hat{\boldsymbol{z}}\;[0]$ & $\boldsymbol{B}\cdot\boldsymbol{\sigma}\;[0]$ & -  \tabularnewline
			\hline
			Ferroelectric\cite{PhysRevB.106.L140505} & $\eta^x$ & $\xi_{\boldsymbol{k}}+\Delta_{SOC}^{(l)}\eta^z\sigma^z\;[1]$ & $\lambda_w(k_x^3-3k_xk_y^2)\eta^z\;[0]$ & $\Delta_z\sigma^z\;[0]$ & -  \tabularnewline
			\hline
			Radial Rashba\cite{PhysRevLett.133.216201} & $\boldsymbol{1}$ & $\xi_{\boldsymbol{k}}\;[0]$ & $\lambda_R\boldsymbol{k}\cdot\boldsymbol{\sigma}\;[0]$ & $\boldsymbol{B}\cdot\boldsymbol{\sigma}\;[0]$ & -  \tabularnewline
			\hline
			$H\left(\eta\phi_{\eta},\mu_{\eta}\right)$\cite{Scammell_2022,RN219} & \multirow{3}*{$\eta^x$} & & & & \tabularnewline
			$\phi_+\neq\phi_-$,$\mu_+=\mu_-$ & ~ & $A_{+}\xi_{\boldsymbol{k}}^e-\mu\;[0]$ & $B_{+}\xi_{\boldsymbol{k}}^o\eta^z\;[0]$ & $A_{-}\xi_{\boldsymbol{k}}^e\eta^z\;[0]$ & $B_{-}\xi_{\boldsymbol{k}}^o\;[0]$  \tabularnewline
			$\phi_+=\phi_-$,$\mu_+\neq\mu_-$ & ~ & $\cos\phi\xi_{\boldsymbol{k}}^e-\mu\;[0]$ & $\sin\phi \xi_{\boldsymbol{k}}^o\eta^z\;[0]$ & $-\delta\mu\eta^z\;[0]$ & -  \tabularnewline
			\hline
			\multirow{2}*{Altermagnet\cite{PhysRevB.110.024503}} & \multirow{2}*{$\boldsymbol{1}$} & \multirow{2}*{$\xi_{\boldsymbol{k}}\;[0]$} & $\boldsymbol{g_k}\cdot\boldsymbol{\sigma}\;[0]$ & $\boldsymbol{N_k}\cdot\boldsymbol{\sigma}\;[0]$ & \multirow{2}*{-}  \tabularnewline
			~ & ~ & & ~ $(\boldsymbol{g_k}=-\boldsymbol{g_{-k}})$ & $(\boldsymbol{N_k}=\boldsymbol{N_{-k}})$ &   \tabularnewline
			\hline
			Weyl and Dirac semimetals\cite{PhysRevB.109.064511} & $\boldsymbol{1}$ & $\xi_{\boldsymbol{k}}\;[0]$ & - & - & $\lambda_0\sin k_i\;[0]$  \tabularnewline
			
		\end{tabular}
	\end{ruledtabular}
	
\end{table*}

The primary task is to determine the relative weights contributed by each permutation. At first glance, one may directly expand Eq.~(\ref{summation}) to obtain
\begin{equation}
	-k_BT\sum_{\boldsymbol{k},i\omega_n}\sum_{n_+,n_-}\mathcal{G}_e^{n_++1}\mathcal{G}_h^{n_-+1}\mathrm{Tr}\left[\left(\delta_{1}+\delta_{2}\right)^{n_+}\left(\delta_{1}-\delta_{2}\right)^{n_-}\right],
\end{equation} 
where the Green's functions are defined as $\mathcal{G}_{e(h)}=[i\omega_n\mp E^{TS,IS}]^{-1}$. For a fixed total power $\{l,n\}$, and using the result of Eq.~(\ref{Chilm}), this expansion can be rewritten as
\begin{align}
	&N_F\sum_{n_+,n_-}\frac{1}{n_+!}\frac{1}{n_-!}\tilde{\chi}_{n_++n_-+2}\mathrm{Tr}\left[\left(\delta_{1}+\delta_{2}\right)^{n_+}\left(\delta_{1}-\delta_{2}\right)^{n_-}\right] \notag \\
	=&N_F\sum_N\frac{2^N}{N!}\tilde{\chi}_{N+2}\mathrm{Tr}\left(\delta_{1}^N\right) \notag \\
	&+\left(\text{terms involving commutators } \left[\delta_{1},\delta_{2}\right]^{m}\right). \label{criterion}
\end{align}
If the commutator terms in parentheses are neglected, all permutations contribute with equal weight, and the criterion reduces to satisfying either of the following inequalities:
\begin{align}
	\sum_P\Pi_1(P)\neq0, \label{Eq1} \\
	\sum_P\Pi_2(P)\neq0, \label{Eq2}
\end{align}
where $\Pi_i(P)$, $i\in\{1,2\}$ are defined in Eqs. (\ref{r1}, \ref{r2}). This simplification is mathematically nontrivial. For instance, expanding \(\mathrm{Tr}[(A+B)^4]\) yields \(\sum_k 4!/[k!(4-k)!]\mathrm{Tr}(A^{4-k}B^k) + \mathrm{Tr}([A,B]^2)\), where the commutator term is not captured by a simple binomial expansion of the trace. 

Nevertheless, we argue that the first term captures the vast majority of the relevant physics and is sufficient to diagnose the emergence of intrinsic SDE. 

On the one hand, Hamiltonian terms with different momentum dependences introduce distinct coefficients when replacing $\delta_{1}$ with $\delta_{2}$. Consequently, any cancellation arising from the commutator terms would require fine-tuning and cannot systematically nullify the first term across all orders in $q$. Therefore, a nonzero result for $\sum_P\Pi(P)\neq0$ identifies a leading algebra-allowed SDE.

On the other hand, we introduce an alternative approximation scheme that clarifies the physical meaning of the commutator terms. Equation~(\ref{summation}) can be recast as
\begin{align}
	-k_BT\sum_{\boldsymbol{k},i\omega_n}\mathrm{Tr}\left[\left(\tilde{\mathcal{G}}_e^{-1}-\delta_1\right)^{-1}\left(\tilde{\mathcal{G}}_h^{-1}-\delta_1\right)^{-1}\right],
\end{align}
where we treat $\delta_{2}$ as a renormalization of the effective Fermi-surface geometry, defining $\tilde{\mathcal{G}}_{e(h)}^{-1}=i\omega_n\mp(E^{TS,IS}+\delta_2)$. Notably, this renormalization does not introduce an energy mismatch between paired electrons. Assuming that $\delta_{2}$ can be diagonalized by a unitary matrix $U_{\boldsymbol{k},\boldsymbol{q}}$, we obtain
\begin{widetext}
	\begin{align}
		&-k_BT\sum_{n_+,n_-}\int d\Omega\, N_F(\Omega) \int_{-\infty}^{+\infty}d\epsilon\sum_{i,j,k,l,m}\delta_{ij}\left[\tilde{\delta}_{1}^{n_+}(\boldsymbol{q},\Omega)\right]_{jk}\delta_{kl}\left[\tilde{\delta}_{1}^{n_-}(\boldsymbol{q},\Omega)\right]_{lm}\delta_{mi} \notag \\
		&\times\left(\frac{1}{i\omega_n-\epsilon-\Delta\epsilon_i}\right)^{n_+}\left(\frac{1}{i\omega_n+\epsilon+\Delta\epsilon_m}\right)^{n_-}\left(\frac{1}{i\omega_n-\epsilon-\Delta\epsilon_k}\right)\left(\frac{1}{i\omega_n+\epsilon+\Delta\epsilon_k}\right),
	\end{align}
\end{widetext}
where $\delta_{ij}$ is the Kronecker delta and $\tilde{\delta}_{1}(\boldsymbol{q},\Omega)=U_{\boldsymbol{k},\boldsymbol{q}}^{\dagger}\delta_{1}U_{\boldsymbol{k},\boldsymbol{q}}$. If we neglect the band-dependent variations of the diagonalized $E^{TS,IS}+\delta_{2}$, i.e., set $\Delta\epsilon\equiv \Delta\epsilon_i$ as a constant, we recover exactly the first term in Eq.~(\ref{criterion}). This indicates that the commutator terms in parentheses of Eq.~(\ref{criterion}) stem from interband pairing contributions, rather than from the Fermi surface shifts or distortions captured by the leading term. Since Fermi surface shifts and distortions constitute the fundamental origin of nonreciprocal physics, it is justified to focus on the first term and adopt Eqs.~(\ref{Eq1}) and (\ref{Eq2}) as the practical criterion for the emergence of intrinsic SDE.

The inequalities \eqref{Eq1} and \eqref{Eq2} constitute one of the central results of this work: intrinsic SDE can emerge whenever either condition is met (the single‑band Fermi‑surface‑shift case discussed in Sec. \ref{sec_alpha} being the sole exception).

We stress that this criterion is formulated in terms of ETR and EI symmetries, not the physical symmetries $\mathcal{T}$ and $\mathcal{P}$, although they coincide in minimal models. Crucially, ETR is defined relative to the specific pairing form. For instance, an Ising SOC term in valley systems preserves $\mathcal{T}$ and breaks $\mathcal{P}$, yet under our ETR definition $\hat{T}=i\eta^x\sigma^y\mathcal{K}$ it is classified as $H^{TS,IS}$, where $(\eta_0,\boldsymbol{\eta})$ span the valley space. Consequently, valley‑dependent and valley‑independent terms typically contribute through distinct channels. As shown in Ref.~\cite{RN290}, SDE cannot originate from Ising SOC and a Zeeman field alone.

\section{Graph-Theoretic construction for nonreciprocal models in $2^N$-DOF \label{sec_graph}}
\subsection{Graph-theoretic mapping of the inequalities}
The same algebraic structure that yields the trace inequalities \eqref{Eq1} and \eqref{Eq2} naturally maps to a graph-theoretic problem when the Hamiltonian is expressed in the Pauli-matrix product basis. In general, the sequence-dependent trace $\Pi_i(P)$ ($i\in\{1,2\}$) is complicated and may be positive, negative, or zero depending on the permutation $P$. Fortunately, real physical systems can typically be described by effective models with $2^N$-DOF. For such models, the Hamiltonian can be expressed in the basis of Pauli matrices products:
\begin{equation}
	H(\boldsymbol{k})=\sum_{\mu_1,...,\mu_N}f_{\mu_1,...,\mu_N}(\boldsymbol{k})\sigma_{\mu_1}\otimes\sigma_{\mu_2}\otimes\cdots\otimes\sigma_{\mu_N}.
\end{equation}
Here, $\sigma_{\mu_j} \in \{I, X, Y, Z\}$ denotes the identity or one of the three Pauli matrices acting on the $j$-th 2-DOF subspace. In this basis, the commutation and anticommutation relations among Pauli matrices ensure that traces for different permutations $\Pi_i(P)$ differ only by an overall sign factor $\epsilon_P=\pm 1$. The criterion $\sum_P\Pi(P)\neq 0$ then separates into two requirements: (i) The trace $\Pi(P_0)\neq 0$ for an arbitrary sequence $P_0$; and (ii) $\sum_P\epsilon_P\neq 0$.

Requirement (i) is satisfied by the appearance of pairs of identical Pauli matrices or the specific group $\{X,Y,Z\}$. We illustrate this using the following models. 
\begin{itemize}
	\item[A.] \textit{Rashba, Radial Rashba, and Altermagnet models:} These share the general Hamiltonian structure $H^{TS,IS}=\xi_{\boldsymbol{k}}$, $H^{TS,IA}=\boldsymbol{g}_{\boldsymbol{k}}\cdot\boldsymbol{\sigma}$, $H^{TA,IS}=\boldsymbol{N}_{\boldsymbol{k}}\cdot\boldsymbol{\sigma}$, and $H^{TA,IA}=0$, with $\boldsymbol{g}_{\boldsymbol{k}}=-\boldsymbol{g}_{-\boldsymbol{k}}$ and $\boldsymbol{N}_{\boldsymbol{k}}=\boldsymbol{N}_{-\boldsymbol{k}}$. The non-trivial condition $\Pi(P)\neq 0$ arises from $\Pi_{1;1}\propto \mathrm{Tr}(X^2)$ and $\Pi_{1;2}\propto \mathrm{Tr}(Y^2)$, which contribute separately.
	\item[B.] \textit{Ferroelectric model:} The relevant matrix terms are $H^{TS,IS}=\xi_{\boldsymbol{k}}+\Delta^{(l)}_{\mathrm{SOC}}\eta^z\sigma^z$, $H^{TS,IA}=\lambda_w(k_x^3-3k_xk_y^2)\eta^z$, and $H^{TA,IS}=\Delta_z\sigma^z$. The non-trivial condition $\Pi(P)\neq 0$ originates from $\mathrm{Tr}[(Z\otimes Z)(I\otimes Z)(Z\otimes I)]$.
\end{itemize}
Crucially, the nonreciprocity of the models~\cite{PhysRevLett.128.037001,RN81,RN98,PhysRevB.106.L140505,PhysRevLett.133.216201,Scammell_2022,RN219,PhysRevB.110.024503,PhysRevB.109.064511} consistently stems from groups of identical Pauli matrices, whose commutativity naturally ensures $\sum_P\epsilon_P\neq 0$.

What if we go beyond these simple cases, where $\Pi(P)\neq 0$ arises from non-commuting terms? Suppose the trace of a matrix sequence $\mathcal{S}_a\mathcal{S}_b\cdots\mathcal{S}_z$ of arbitrary length is nonzero, where $\mathcal{S}\equiv\sigma_{\mu_1}\otimes\sigma_{\mu_2}\otimes\cdots\otimes\sigma_{\mu_N}$. The cyclic property of the trace, $\mathrm{Tr}(\mathcal{S}_a\mathcal{S}_b\cdots\mathcal{S}_z)=\mathrm{Tr}(\mathcal{S}_b\cdots\mathcal{S}_z\mathcal{S}_a)$, implies that each term must exhibit an even number of anticommutation relations with the other terms. Moreover, every permutation $\mathcal{S}_a\mathcal{S}_b\cdots\mathcal{S}_z$ has a unique reciprocal counterpart $\mathcal{S}_z\cdots\mathcal{S}_b\mathcal{S}_a$, which implies that the total number of anticommutation relations is also even. Taking these two constraints together, we can represent the models using graphs. 

The rule is as follows: vertices represent distinct matrices appearing in the Hamiltonian terms, and edges represent anticommutation relations between the corresponding matrices. Here, "distinct matrices" means that different vertices correspond to different Pauli products; the identity matrix $\boldsymbol{1}$ commutes with all matrices and therefore never participates in anticommutation relations, so it does not appear in the graph. Repeated occurrences of the same matrix are excluded from the graph for a different reason: they already provide a trivial mechanism for satisfying the two requirements without any need for graph-theoretic analysis. Specifically, two identical matrices share the same matrix part under the $\hat{T}$ transformation. By endowing them with scalar functions of opposite $\bm{k}$-parity, their EI symmetry can be made opposite, while the full ETR transformation $\hat{T}\hat{I}$ then assigns them opposite ETR parity as well. They therefore belong to $H^{TS,IA}$ and $H^{TA,IS}$ [the channel of Eq.~\eqref{Eq1}], or to $H^{TS,IS}$ and $H^{TA,IA}$ [the channel of Eq.~\eqref{Eq2}]. In either case, the nonvanishing of the trace is directly guaranteed by $\mathcal{S}^2=\boldsymbol{1}$. Such repeated-matrix mechanisms are simple and do not require graph-theoretic analysis.

\begin{figure}[htbp]
	\begin{center}
		\includegraphics[width=1.0\columnwidth]{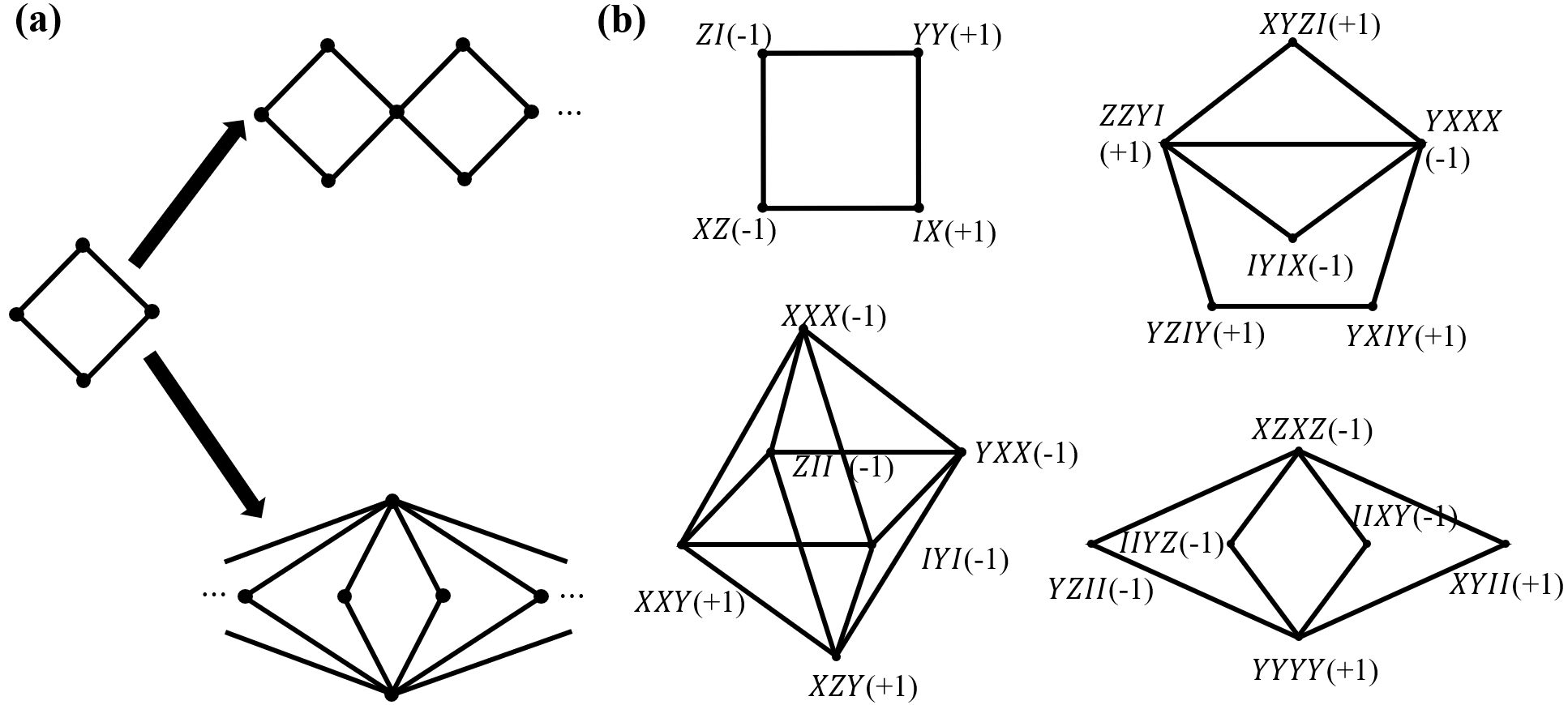}
	\end{center}
	\caption{Graph-theoretic representation of nonreciprocal models. (a) Complex nonreciprocal models can be constructed by combining elementary single cycles, illustrating the inheritance of nonreciprocity from simpler building blocks. (b) Correspondence between explicit Hamiltonian models and their associated graphs. Notably, the model corresponding to the rectangular cycle has been reported in Ref.~\cite{fracassi2025intrinsictunablesuperconductingdiode}. }\label{graph}
\end{figure}

The resulting graphs possess the following characteristics:
\begin{enumerate}
	\item[a.] Every vertex has even degree (i.e., every matrix has even number of anticommutation relations).
	\item[b.] The total number of edges is even.
\end{enumerate}
We illustrate several examples in Fig.~\ref{graph}. 

In this manner, the problem of diagnosing nonreciprocal models is reduced to a purely graph-theoretic question: evaluating the permutation sum $K\equiv\sum_P\epsilon_P$ for a given graph.

\subsection{Analytic evaluation of cycle contributions}
The simplest graphs satisfying the above characteristics are single cycles. Consider a single cycle $C_n$ consisting of $n$ vertices and $n$ edges. The vertices ($v_i$, $i\in\{1,...,n\}$) represent products of Pauli matrices in the $2^N$-DOF system, and each edge connecting $v_i$ and $v_{i+1}$ (with the convention $v_{n+1}=v_1$) represents an anticommutation relation between the corresponding matrices. An arbitrary permutation $P$ of the matrix sequence acquires a sign $\epsilon_P=\pm 1$ relative to the reference ordering, determined by the number of anticommuting exchanges. The sum over all $n!$ permutations defines the quantity
\begin{equation}
	K_n\equiv\sum_{P}\epsilon_P.
\end{equation}

The cyclic symmetry of the ring substantially simplifies the calculation. Under a cyclic shift of the entire sequence, the trace remains invariant by its cyclic property, and the sign $\epsilon_P$ is also unchanged. This is because each step of the cyclic shift moves the first matrix past its two neighbors---the second matrix and the last matrix---via anticommutation, contributing a factor of $(-1)^2 = +1$. One may therefore fix a particular matrix at the first position, leaving the remaining $n-1$ matrices to be freely permuted. This effectively breaks the cycle at that vertex, reducing it to a chain. The resulting relation is
\begin{equation}
	K_n=nT_{n-1},
\end{equation}
where $T_{n-1}$ is the permutation sum for a chain of $n-1$ vertices.

For the chain, consider an arbitrary permutation and its reverse (the sequence read backwards). Since anticommutation relations only exist between adjacent vertices in the chain, reversing the sequence swaps the two vertices connected by each edge, contributing a factor of $(-1)^{n-2}$ from the $n-2$ edges. When $n-2$ is odd, a permutation and its reverse carry opposite signs and cancel pairwise; hence $T_{2l}=0$, and only even $n$ can yield a nonvanishing result.

To evaluate the chain sum for even $n$, we single out a specific vertex and place it at the first position. Suppose this vertex is the $k$-th vertex in the normal ordering of the chain. Removing it breaks the chain into a left segment of $k-1$ vertices and a right segment of $n-k-1$ vertices. Since anticommutation relations only exist between neighboring vertices, moving the selected vertex across the left segment involves only a single anticommutation with its immediate neighbor, introducing a factor of $-1$. The remaining vertices are distributed among the two segments in all possible ways, with the ordering within each segment preserved. This yields the recursion relation
\begin{equation}
	T_{n-1}=-\sum_{k=1}^{n-1}\binom{n-2}{k-1}T_{k-1}T_{n-k-1},
\end{equation}
with the convention $T_0=1$. 

Using $T_{2l}=0$ and applying the substitution $T_{2l+1}=(-1)^lT_{l+1}^*$, the recursion reduces to
\begin{equation}
	T_{l+1}^*=\sum_{k=1}^{l}\binom{2l}{2k-1}T_k^*T_{l-k+1}^*,
\end{equation}
which is precisely the defining equation for the tangent numbers $T_k^*$, with the initial condition $T_1^*=1$.

Assembling these results, the permutation sum for a single cycle can be obtained in closed form. For an odd cycle ($n=2m+1$),
\begin{equation}
	K_{2m+1}=0.
\end{equation}
For an even cycle ($n=2m$),
\begin{equation}
	K_{2m}=2^{2m}\left(2^{2m}-1\right)B_{2m}\neq0,
\end{equation}
where $B_k$ are the Bernoulli numbers ($B_2=1/6$, $B_4=-1/30$, $B_6=1/42$, ...). 

The appearance of Bernoulli numbers indicates that the condition for nonreciprocity in this algebraic setting is governed by combinatorial rules, rather than by the details of the band parameters or matrix representations. This result is the cornerstone of the graph-theoretic construction developed below.

\subsection{Construction for nonreciprocal models}
We now determine the characteristics of complex graphs built from the elementary single cycles analyzed above. Consider a graph in which every vertex has even degree. Starting from an arbitrary vertex, traverse an edge to a neighboring vertex, then exit via a different edge to the next vertex, and continue this procedure without repeating any vertex (except the starting point) or edge. The path eventually returns to the initial vertex, forming a closed loop---a single cycle. After peeling off this cycle from the original graph, the remaining subgraph still satisfies the condition that every vertex has even degree, and therefore belongs to the same general class we started with. This peeling procedure can be repeated until all edges are exhausted. Consequently, any graph whose vertices all have even degree can be decomposed into a set of single cycles that share only common vertices and no common edges. Single cycles thus constitute the fundamental building blocks of such graphs. Whether the full graph inherits the nonreciprocity properties of the individual cycles, however, requires further analysis.

The way in which cycles are interconnected determines whether the nonreciprocity of the full graph follows directly from that of its constituent single cycles. When two subgraphs share only a single common vertex, the permutation signs associated with each subgraph remain independent, and the total permutation sum equals the product of the permutation sums of the two subgraphs, multiplied by an appropriate combinatorial factor accounting for the arrangements of commuting vertices belonging to different subgraphs. In this case, if the permutation sum of either subgraph vanishes (as for an odd cycle with $K_{2m+1}=0$), the permutation sum of the entire graph vanishes as well. Hence, a complex graph constructed solely from single cycles sharing single common vertices fully inherits the properties of its building blocks.

When two cycles share a pair of common vertices, the situation is more subtle. Sharing a pair of vertices while retaining independent edges introduces a correlation between the permutation signs of the two cycles, so that the total permutation sum no longer factorizes into the product of the individual contributions. A graph composed purely of even cycles can still inherit the nonreciprocity of its constituents. Notably, however, a pair of odd cycles sharing two common vertices can combine to yield a graph with a nonvanishing total permutation sum. Odd cycles are therefore not strictly excluded from contributing to nonreciprocity in this more general setting.

From the perspective of nonreciprocity inheritance, not all complex graphs can be reduced to simple assemblies of single cycles. Certain complex graphs---such as the octahedron graph---constitute irreducible building blocks on an equal footing with single cycles, whose permutation sums must be computed independently and cannot be derived from single-cycle results (the octahedron graph yields $|K|=48\neq 0$). The complete graph-theoretic construction therefore comprises both single cycles and other irreducible complex building blocks, combined through shared vertices. Any graph with a nonvanishing permutation sum, whether decomposable into single cycles or not, can serve as a graph-theoretic blueprint for a nonreciprocal model.

This graph-theoretic framework provides a systematic graphical method for generating nonreciprocal models. The complete construction procedure is summarized as follows:
\begin{enumerate}
	\item \textbf{Construct the graph.} Draw a connected or disconnected graph in which every vertex has even degree and the total number of edges is even. The graph may be assembled from single cycles sharing one or two common vertices, or may directly adopt an irreducible building block (such as the octahedron graph), or a combination of irreducible blocks connected through shared vertices. Each edge of the graph represents an anticommutation relation between the matrices assigned to its two endpoints.
	\item \textbf{Assign matrices.} Choose a Pauli-matrix product space of $2^N$ dimensions. Assign to each vertex of the graph a nontrivial (i.e., not the identity matrix $\boldsymbol{1}$) basis matrix $\mathcal{S}_a$, with distinct matrices assigned to distinct vertices. The edges of the graph are in one-to-one correspondence with the anticommutation relations among the assigned matrices. In addition, ensure that the product of all vertex matrices is proportional to the identity.
	\item \textbf{Determine the parity of each term under the $\hat{T}$ transformation.} By convention, select the matrix part of the ETR operation as $\hat{T}_0=iY\otimes I\otimes \cdots\otimes I \,\mathcal{K}$. Compute the eigenvalue of each vertex matrix under the pure matrix transformation $\hat{T}_0$.
	\item \textbf{Assign scalar functions.} Generate the Hamiltonian terms by attaching a specific $\boldsymbol{k}$-dependent function $f_a(\boldsymbol{k})$ (such as $k_x$, $k_y$, $k^2$, or a constant) to each vertex $\mathcal{S}_a$. Note that for a general vertex term $f_a(\boldsymbol{k})\mathcal{S}_a$, the ETR symmetry classification is determined by the product of the parity of the scalar function $f_a(\boldsymbol{k})$ and the eigenvalue of the matrix part under $\hat{T}_0$, while the EI symmetry classification is determined solely by the parity of $f_a(\boldsymbol{k})$. The sector assignments must ensure that $H^{TS,IA}$ and $H^{TA,IS}$ each appear an odd number of times (typically once each) to satisfy Eq.~\eqref{Eq1}, or that $H^{TS,IS}$ and $H^{TA,IA}$ each appear an odd number of times to satisfy Eq.~\eqref{Eq2}.
\end{enumerate}
For a Hamiltonian constructed according to this procedure, the matrix assignment in step 2 and the sector assignment in step 4 together guarantee condition (i) (that a product proportional to the identity matrix exists for some permutation, yielding a nonzero trace), while the graph topology guarantees condition (ii) (that the permutation sum $K=\sum_P\epsilon_P\neq 0$, so contributions from different permutations do not cancel). The trace inequalities~\eqref{Eq1} or~\eqref{Eq2} are then automatically satisfied, and the model necessarily supports intrinsic SDE. Crucially, the core algebraic structure of the model is already fixed in the first two steps; the remaining two steps merely translate this structure into a concrete Hamiltonian. A given graph therefore corresponds to an entire family of equivalent nonreciprocal models, reflecting a degree of arbitrariness in this correspondence.

To summarize, we have established the other main result of this work: a generating procedure for nonreciprocal $2^N$-DOF models, consisting of drawing graphs and assigning explicit matrices to the vertices. We emphasize that each graph corresponds to a class of mathematically equivalent models, and the nonreciprocity stems from the topological properties of the graph rather than the specific physical meaning of the individual vertices.

\section{Numerical verifacation of the graph-theoretic constructed 1D models \label{sec_numerical}} 
Based on the graph-theoretic construction presented, nonreciprocal models can be systematically generated by drawing graphs and assigning explicit matrices to their vertices. In this section, we list several representative graphs and provide one corresponding microscopic model for each. Throughout, we adopt the ETR operator $\hat{T}_0 = iY \otimes I \otimes \cdots \otimes I \,\mathcal{K}$. Matrices with even parity under $\hat{T}_0$ are assigned to $H^{TS,IS}$ or $H^{TA,IA}$, and those with odd parity to $H^{TS,IA}$ or $H^{TA,IS}$. For simplicity, all examples below take the nonreciprocal contribution to arise from $H^{TS,IA}$ and $H^{TA,IS}$. The direct product symbol $\otimes$ is omitted throughout. The presence or absence of nonreciprocity is verified by plotting the band-energy difference $E(k)-E(-k)$, which proves to be robust under continuous tuning of the model parameters.

\subsection{Single rectangular cycle} 
This graph [as shown in Fig. \ref{graph} (b)] corresponds to a model reported in Ref.~\cite{fracassi2025intrinsictunablesuperconductingdiode}, which we summarize as the following $2^2$-DOF Hamiltonian:
\begin{equation}
	H_{\text{r}}=k(XZ)+U_z(ZI)+\tau_p(IX)+\tau_f(YY).
\end{equation}
The terms $k(XZ)$ and $U_z(ZI)$ have odd parity under $\hat{T}_0$ and are assigned to $H^{TS,IA}$ and $H^{TA,IS}$, respectively. Our framework immediately reveals why a nonzero $\tau_p$ is essential for the emergence of the SDE.

\subsection{Two rectangles sharing a single common vertex} 
\begin{figure}[htp]
	\centering
	\includegraphics[width=1.0\columnwidth]{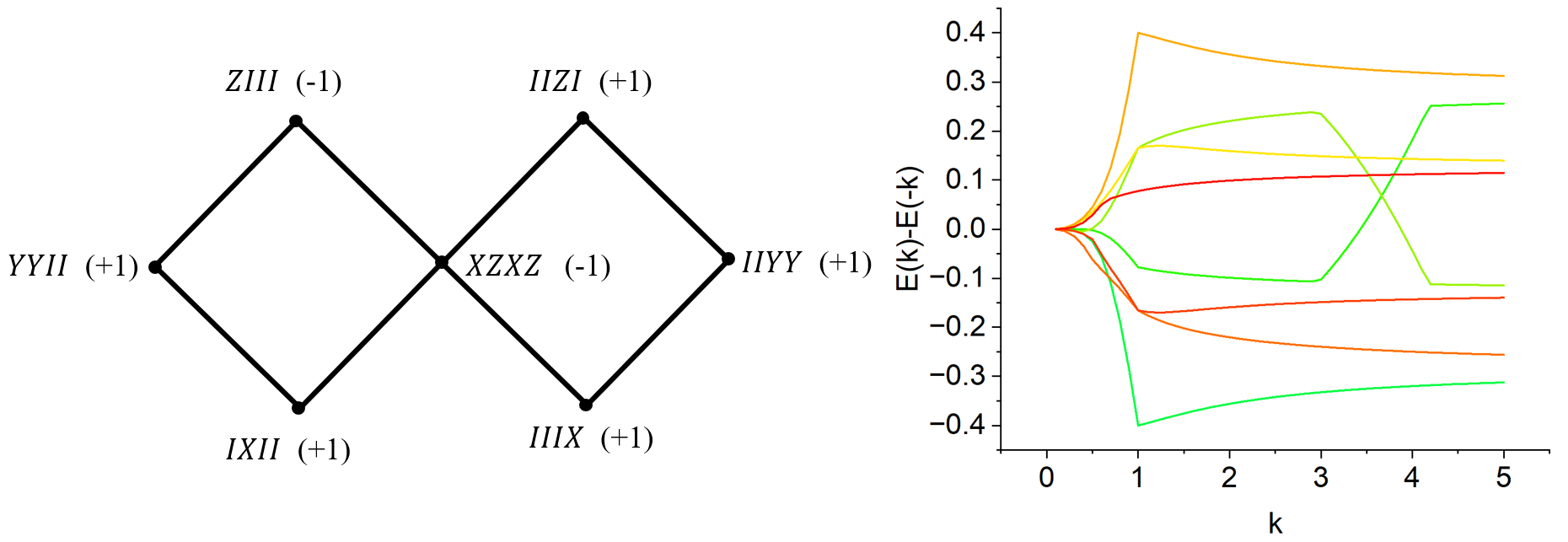}
	\caption{ The graph of two rectangles sharing a single common vertex and the nonreciprocity of its corresponding model. \label{double_square}}
\end{figure}

The graph and matrix assignment are shown in Fig.~\ref{double_square}. A $2^4$-DOF realization is
\begin{align}
	H_{\text{r-r}}=&t_0k^2(IIII)+\alpha k(ZIII)+\beta(XZXZ) \notag \\
	&+t_1k^2(YYII)+t_2k^2(IXII)+t_3 k^2(IIZI) \notag \\
	&+t_4 k^2(IIYY)+t_5 k^2(IIIX).
\end{align}
The terms $\alpha k (ZIII)$ and $\beta (XZXZ)$ are assigned to $H^{TS,IA}$ and $H^{TA,IS}$.

\subsection{Two rectangles sharing a pair of common vertices}
\begin{figure}[htp]
	\centering
	\includegraphics[width=1.0\columnwidth]{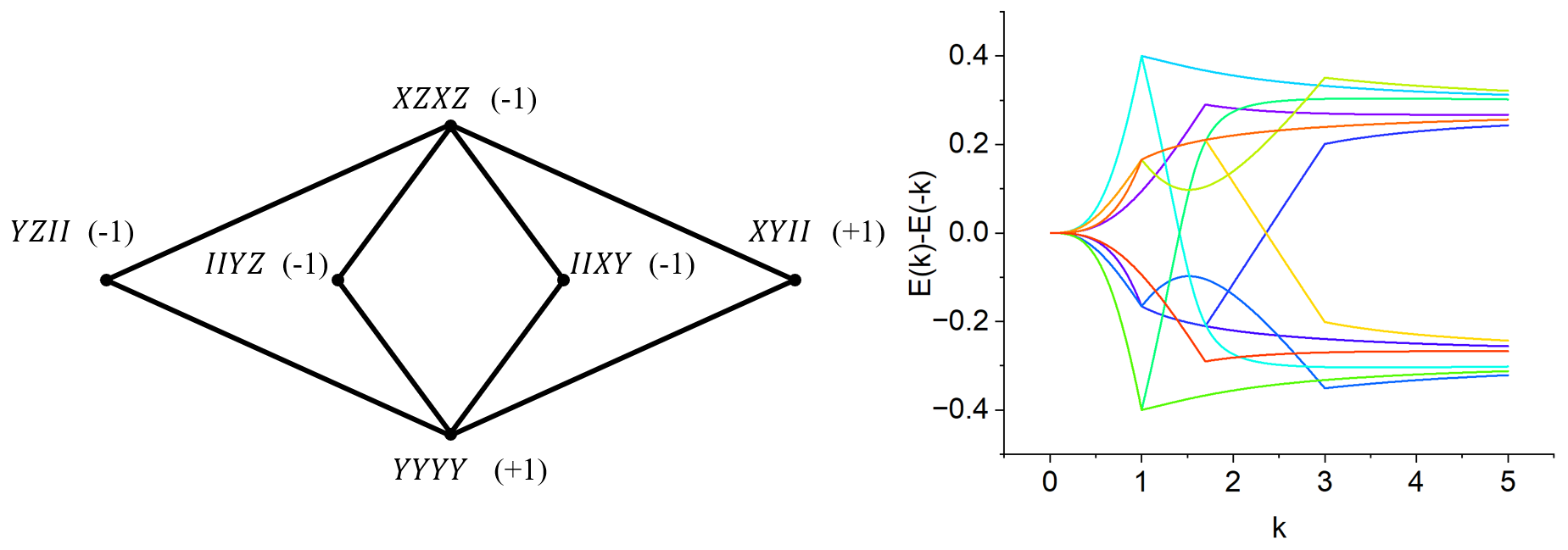}
	\caption{ The graph of two rectangles sharing a pair of common vertices and the nonreciprocity of its corresponding model. \label{two_square}}
\end{figure}

The graph and matrix assignment are shown in Fig.~\ref{two_square}. A $2^4$-DOF realization is
\begin{align}
	&H_{\text{rr}}=t_0 k^2(IIII)+\alpha k(XZXZ)+\beta_1(YZII) \notag \\
	&+\beta_2(IIYZ)+\beta_3(IIXY)+t_1 k^2(XYII)+t_2 k^2(YYYY).
\end{align}
Among the four vertices with odd parity under $\hat{T}_0$, one is assigned to $H^{TS,IA}$ and the remaining three to $H^{TA,IS}$.

\subsection{Octahedron}
\begin{figure}[htp]
	\centering
	\includegraphics[width=1.0\columnwidth]{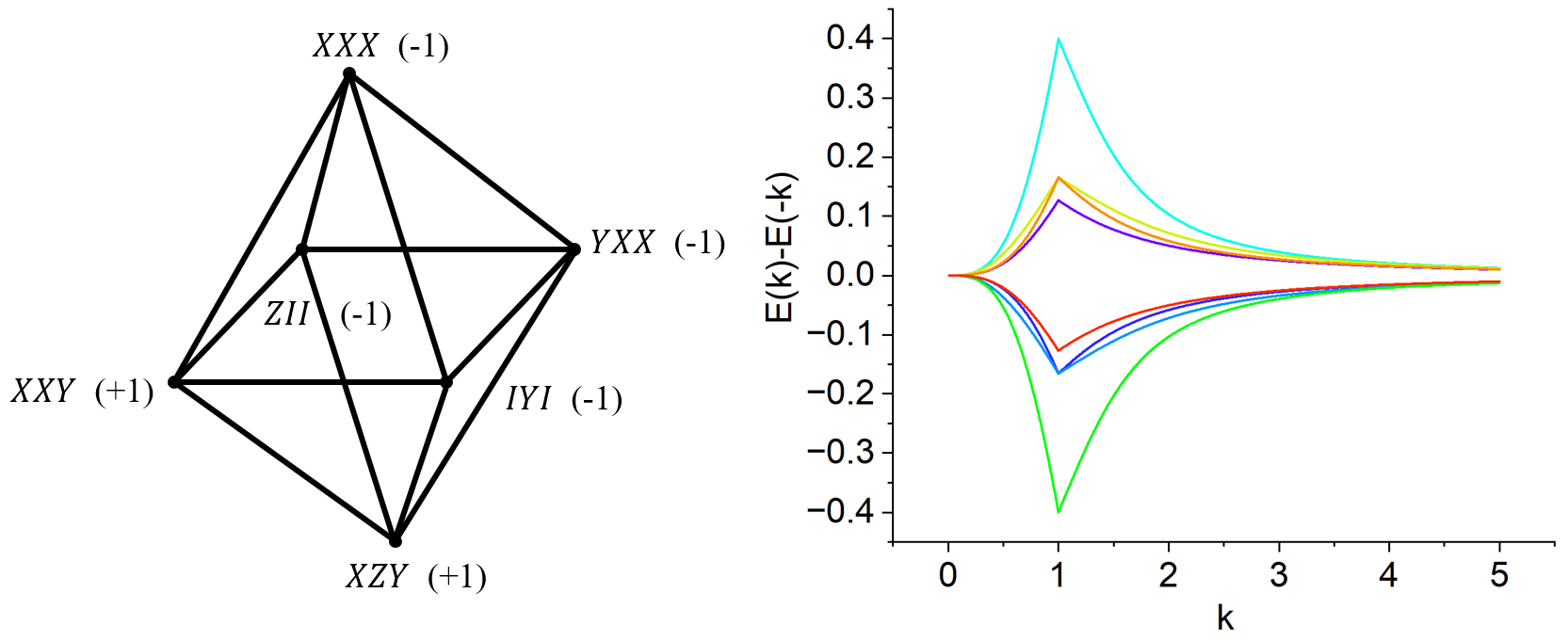}
	\caption{ The octahedron graph and the nonreciprocity of its corresponding model. \label{Octahedron}}
\end{figure}

The octahedron graph (Fig. \ref{Octahedron}) constitutes an irreducible building block beyond single cycles, with permutation sum $|K|=48$. A $2^3$-DOF realization is
\begin{align}
	H_{\text{o}}=&t_0 k^2(III)+\alpha k(XXX)+\beta_1(ZII)+\beta_2(YXX) \notag \\
	&+\beta_3(IYI)+t_1 k^2(XXY)+t_2 k^2(XZY).
\end{align}
The four vertices with odd parity under $\hat{T}_0$ are assigned to one $H^{TS,IA}$ and three $H^{TA,IS}$.

\subsection{Triangle inscribed within a pentagon}
\begin{figure}[htp]
	\centering
	\includegraphics[width=1.0\columnwidth]{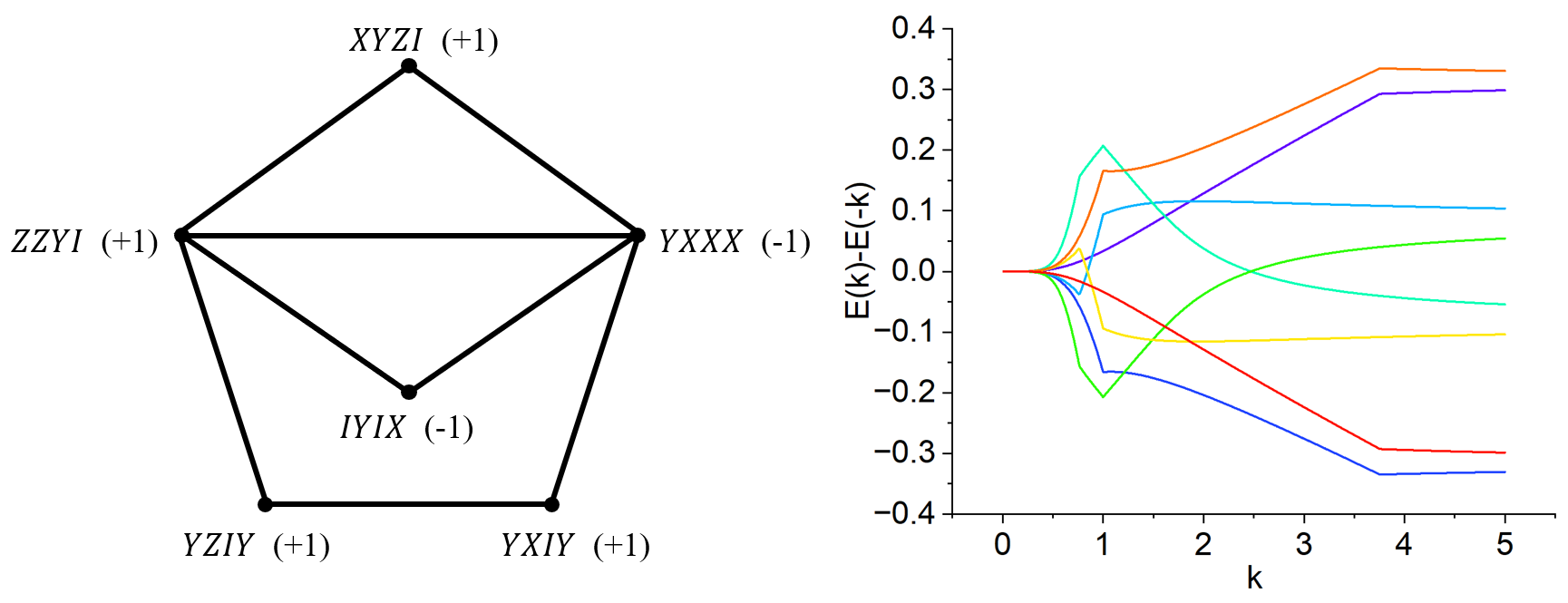}
	\caption{ The graph of a triangle inscribed within a pentagon and and the nonreciprocity of its corresponding model. \label{pentagon}}
\end{figure}

This graph (Fig. \ref{pentagon}) consists of two odd cycles sharing a pair of common vertices, yielding a nonvanishing permutation sum $|K|=48$. A $2^4$-DOF realization is
\begin{align}
	H_{\text{p}}=&t_0 k^2(IIII)+\alpha k(YXXX)+\beta(IYIX) \notag \\
	&+t_1 k^2(XYZI)+t_2k^2(ZZYI)+t_3 k^2(YZIY) \notag \\
	&+t_4 k^2(YXIY).
\end{align}
The terms $\alpha k(YXXX)$ and $\beta(IYIX)$ are assigned to $H^{TS,IA}$ and $H^{TA,IS}$.

\subsection{Graphs composed of odd single cycles} 
\begin{figure}[htp]
\centering
\includegraphics[width=1.0\columnwidth]{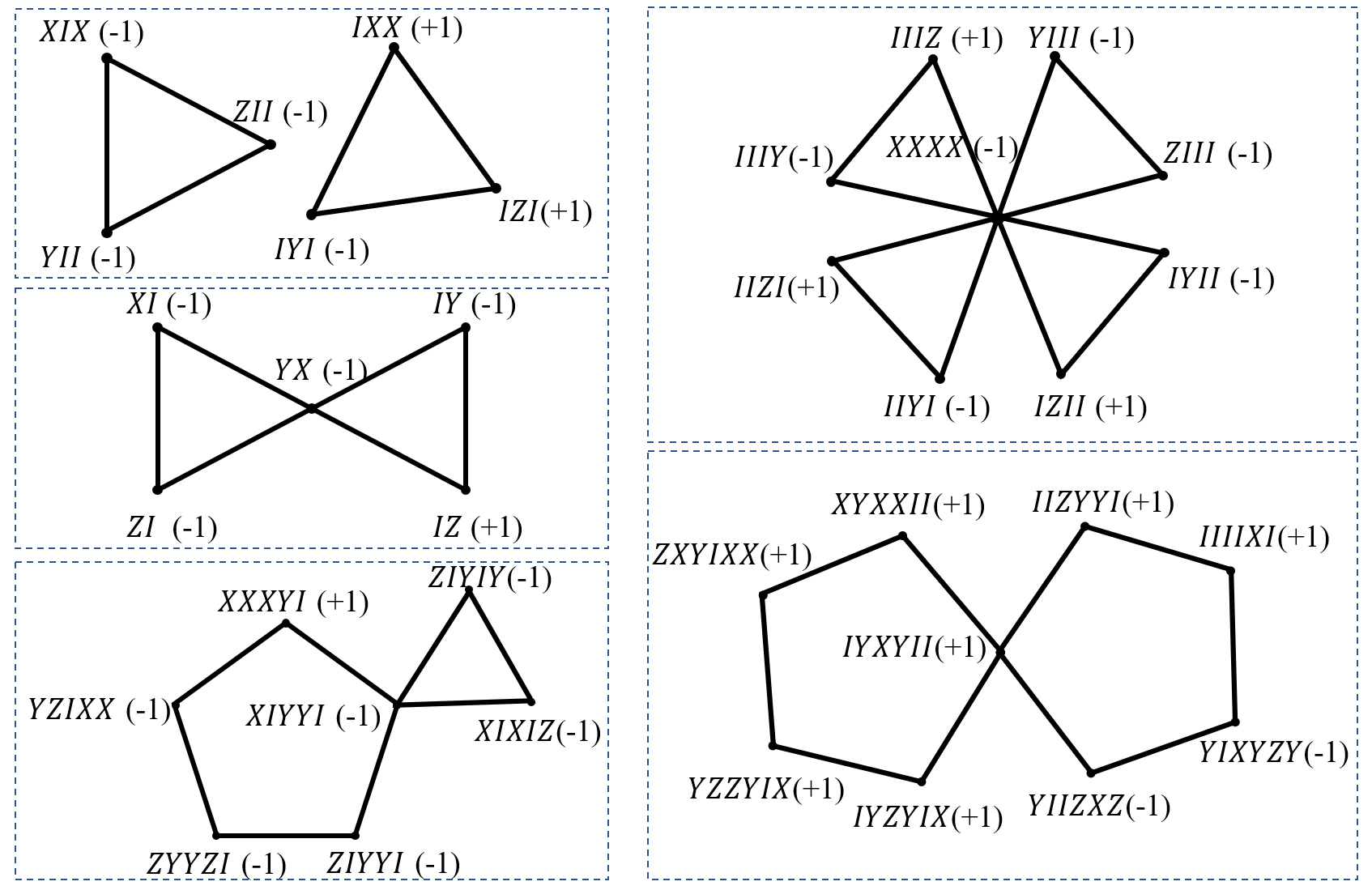}
\caption{ The graphs composed of odd single cycles. Each dashed box contains a total graph. \label{odd_cycles}}
\end{figure}

In addition to the graphs that yield a nonzero permutation sum $K \equiv \sum_P \epsilon_P$ and consequently exhibit nonreciprocal band structures, one can also construct models based on graphs built from odd single cycles. According to our earlier analysis, such graphs give $K = 0$ and should therefore display symmetric bands. Several examples are illustrated in Fig.~\ref{odd_cycles}, and the numerical results confirm that all of them satisfy $E(k) = E(-k)$, i.e., nonreciprocity is absent. This provides direct numerical confirmation of the graph-theoretic criterion. 

\section{Generalization of the criterion to effective bands \label{sec_effective}}
\begin{figure}[htbp]
	\begin{center}
		\includegraphics[width=1.0\columnwidth]{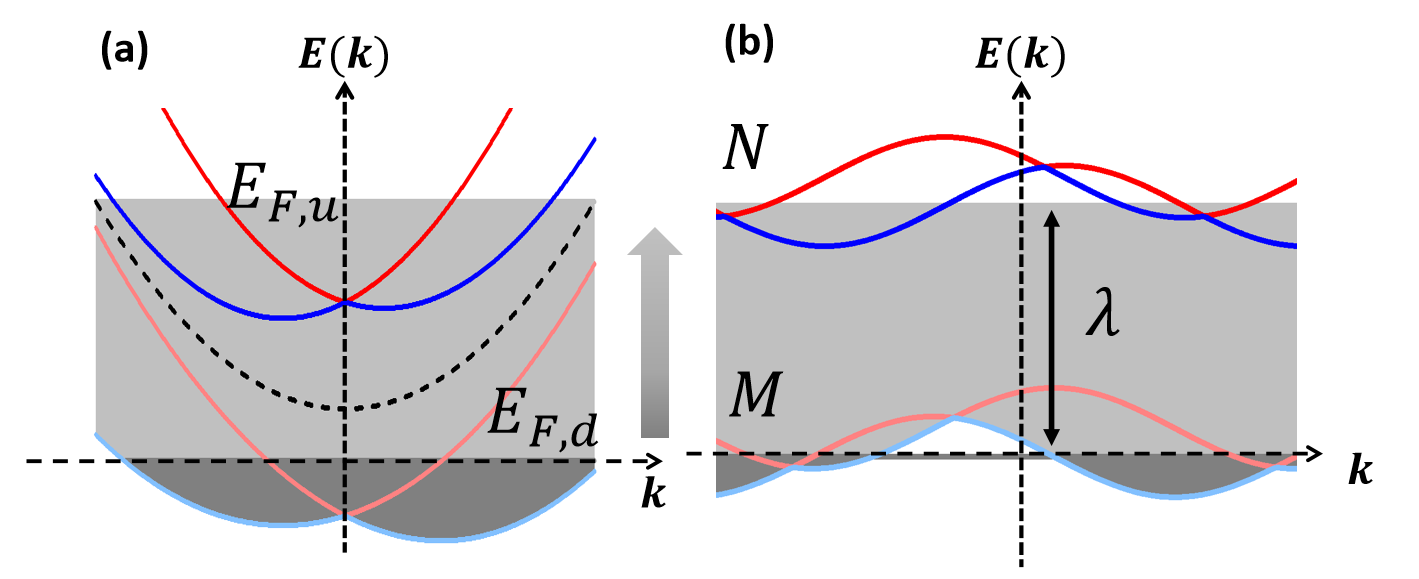}
	\end{center}
	\caption{Schematic for multi bands. (a) Two different Fermi surfaces for the unconventional Rashba band with in-plane external field. The below surface $E_{F,d}$ denotes the unconventional Rashba case, while the above one $E_{F,u}$ denotes the corresponding well-defined model. (b) The bands for the model divided by $M$ and $N$ subspaces. We use two bands on each Fermi surface to represent arbitrary bands in subsets $M$ and $N$.  \label{fig2}}
\end{figure}

The preceding criterion was derived under the assumption of well-defined bands, where all Fermi surfaces intersect every band. However, in many realistic multi-orbital or multi-band systems, such as the unconventional Rashba model (URM) \cite{PhysRevB.109.195419,PhysRevB.110.134517}, the Fermi energy may intersect only a subset of the bands [as shown in Fig. \ref{fig2} (a)]. In such cases of effective bands, the applicability of our trace-based criterion requires careful examination.

To demonstrate that the underlying band asymmetry still governs nonreciprocity, we analyze the URM as a representative example. In this model, the chemical potential lies within the spin-orbit gap, isolating two active bands. Despite the absence of the upper two bands at the Fermi surface, an in-plane Zeeman field $\boldsymbol{B}$ induces a net shift of the Fermi pockets. Crucially, in contrast to the conventional Rashba model where nonreciprocity stems from the difference between opposite band shifts, the SDE in URM originates from the coherent sum of Fermi surface displacements along the same direction (a detailed calculation is provided in Appendix \ref{simplified}). This confirms that the qualitative link between band asymmetry and nonreciprocity remains robust beyond the well-defined limit.

Building on the insights from previous sections, we now generalize the SDE criterion to effective bands. While raising the chemical potential can convert effective bands into well-defined ones, this approach fails for models with true band gaps (e.g., insulators or flat bands). Such cases can be addressed by adding a term proportional the identity matrix  (e.g., a parabolic dispersion $tk^2$) to eliminate the gap. We establish two key principles: (a) If SDE exists in well-defined bands, it must also appear in their corresponding effective bands; (b) Effective bands exhibiting SDE cannot correspond to well-defined bands lacking SDE. Together, these principles ensure a consistent emergence of SDE between well-defined bands and their effective counterparts. 

We consider a multi-band model described by bilinear electron field operators with multi-DOF, as illustrated in Fig. \ref{fig2}(b). The momentum-space Hamiltonian adopts the block-matrix form:
\begin{equation}
	H_{\text{multi}}=\begin{pmatrix}
		\hat{H}_N & \hat{V} \\ \hat{V}^{\dagger} & \hat{H}_M
	\end{pmatrix}.
\end{equation}
To explicitly model the bandgap $\lambda$ between upper and lower bands, we partition the Hilbert space into $M$ and $N$ subspaces. Applying a unitary transformation $\hat{U}$, diagonalizes each subspace individually yields:
\begin{equation}
	\hat{U}^{\dagger}H_{\text{multi}}\hat{U}=\begin{pmatrix}
		\hat{E}_N & U_N^{\dagger}\hat{V}U_M \\ U_M^{\dagger}\hat{V}^{\dagger}U_N & \hat{E}_M
	\end{pmatrix}=\begin{pmatrix}
		\hat{E}_N & \tilde{V} \\ \tilde{V}^{\dagger} & \hat{E}_M
	\end{pmatrix}, 
\end{equation}
where $\hat{E}_M$ and $\hat{E}_N$ are diagonal matrices. The substantial inter-subspace gap $\lambda\gg|\hat{V}|$, ensures $\hat{V}$ acts as a perturbation. Thus, $\hat{E}_N$ and $\hat{E}_M$ provide the zeroth-order band approximation. As established previously, SDE arises from band asymmetries. We therefore decompose the eigenenergies for each subspace $Z\in\{M,N\}$ as $\hat{E}_Z=\hat{E}_Z^S+\hat{E}_Z^A$, where $\hat{E}_Z^S$ is symmetric under $\boldsymbol{k}\to-\boldsymbol{k}$ while $\hat{E}_Z^A$ being antisymmetric. The inter-subspace coupling $\tilde{V}=\tilde{V}^S+\tilde{V}^A$ may also contribute to SDE. Second-order perturbation theory yields the energy correction: $E_n=E_n^{(0)}-\sum_m|\langle m|\tilde{V}|n\rangle|^2/[E_m^{(0)}-E_n^{(0)}]$, where $E_n^{(0)}(E_m^{(0)})$ denotes unperturbed energies in subspace $N$($M$). Assuming $N$ represents upper bands, $E_m^{(0)}-E_n^{(0)}\approx-\lambda$, two implications follow: (1) The band repulsion shifts the $M$- and $N$-bands in opposite directions; (2) The net SDE contribution requires nonzero asymmetry ($\hat{E}_Z^A$ or $\tilde{V}^A$) in both subspaces. Thus, SDE in the full Hilbert space necessitates intrinsic SDE in each subspace when they remain coupled, which establishing our first principle (a). 

The Fermi surface asymmetry is quantified by the antisymmetric energy component at $\boldsymbol{k}_{F,n}$,
\begin{align}
	&E_n^A|_{\boldsymbol{k}_{F,n}}=\left[E_n^{A(0)} \right. \notag \\
	&\left.+\frac{1}{\lambda}\sum_{m}\left(\langle m|\tilde{V}^S|n\rangle\langle n|\tilde{V}^A|m\rangle+c.c.\right)\right]_{\boldsymbol{k}_{F,n}}.
\end{align}
For bands in subspace $M$, apply $m \leftrightarrow n$ and $\lambda \to -\lambda$. The nonreciprocal contribution to the linear order of $\boldsymbol{q}$ from well-defined bands is $|\sum_{m}N_F^mv_F^mE_m^A|_{k_{F,m}}+\sum_{n}N_F^nv_F^nE_n^A|_{k_{F,n}}|$. Here, the velocity for the band $z$ is $\boldsymbol{v}_F^z=\nabla_{\boldsymbol{k}}E_z^S$, and the DOS $N_F^z\propto k_{F,z}/v_F^z$. The bandgap $\lambda$ enforces distinct Fermi circles, yielding different $k_F$ for $M$ and $N$ subspaces. Therefore, even if $E^A$ itself is independent of $k_F$, the overall SDE efficiency cannot cancel out. This demonstrates that SDE in effective bands implies SDE in the associated well-defined bands—establishing principle (b).
The above discussions imply that the qualitative SDE criterion can be safely and consistently extended from well-defined bands to their effective-band counterparts.

\section{Discussions \label{sec_conclu}}
We note that our criterion fundamentally relies on the condition $\hat{T}^2 = \pm 1$, which guarantees a consistent symmetry characterization under ETR. This requirement is fulfilled by all unitary pairing states, including spin-singlet and unitary spin-triplet pairings, but excludes non-unitary states for which a well-defined $\hat{T}$ transformation does not exist. Moreover, although the SDE has been shown to survive in strongly disordered Rashba systems~\cite{PhysRevLett.128.177001}, the present criterion is derived within the standard Ginzburg–Landau framework in the clean limit. It is likewise inapplicable to regimes beyond the mean-field approximation where superconducting fluctuations become significant. The extension of this criterion to the strong-coupling regime remains an open question and a worthwhile direction for future work.

Despite these limitations, the inequalities in Eqs.~\eqref{Eq1} and~\eqref{Eq2} furnish a systematic diagnostic tool for the emergence of intrinsic SDE and substantially advance the fundamental understanding of nonreciprocal superconductivity. The criterion is particularly valuable in multiband or multi-DOF systems~\cite{Yerin_2023,PhysRevB.110.094515,PhysRevB.110.134517,PhysRevB.111.L100506,PhysRevB.111.144513,PhysRevResearch.7.L012010}, where it offers a computationally efficient alternative to numerically intensive approaches, thereby enabling the rapid prediction and rational design of nonreciprocal behavior in novel materials. 

Beyond its immediate application to SDE, the graph-theoretic construction unveiled by the same algebraic structure carries a broader conceptual implication. It demonstrates that the condition for nonreciprocity in these systems is not merely a matter of symmetry classification, but is governed by an underlying combinatorial rule that can be visualized and manipulated as a graph. This offers a complementary perspective: instead of asking which symmetries are broken, one may ask what graph is formed by the anticommuting terms, providing a fresh approach to the long-standing problem of designing nonreciprocal systems.

Together with the diagnostic criterion, the graph-theoretic construction thus provides a complete and coherent methodology: from diagnosing the presence of SDE in a given model, to systematically generating new models with built-in nonreciprocity. 

We note that the restriction to $2^N$-DOF systems is not a fundamental limitation of the graph-theoretic approach. Any system with a finite number of DOF can be formally embedded into a larger $2^N$-DOF space. Algebraically, this is achieved by zero-padding the Hamiltonian matrix in the additional dimensions; physically, it is equivalent to adding decoupled, symmetric bands that do not contribute to nonreciprocal responses. When expanded in the Pauli-matrix product basis, the anticommutation relations among the resulting matrix components necessarily map to a graph. This observation suggests that the connection between nonreciprocal physics and graph theory may be deeper than the specific construction presented here, pointing towards an intrinsic, algebraic-geometric origin of nonreciprocal transport that warrants further investigation.

We anticipate that this framework will stimulate further theoretical investigations into the deeper connections between nonreciprocal physics and discrete mathematics, and may ultimately find analogues beyond the realm of superconductivity.

\begin{acknowledgments}
We thank Daniel Shaffer for helpful discussions. We thank Ming-Wei Wu and Song-Bo Zhang for their encouragement. This work was financially supported by the National Key R\&D Program of China (Grants No. 2024YFA1613200 and No. 2022YFA1403200), National Natural Science Foundation of
China (Grants No. 92265104, No. 12022413), the Basic Research Program of the Chinese
Academy of Sciences Based on Major Scientific Infrastructures (Grant No. JZHKYPT-2021-08), the CASHIPS Director’s Fund (Grant No. BJPY2023A09), Anhui Provincial Major S\&T Project(s202305a12020005), and
the High Magnetic Field Laboratory of Anhui Province under Contract No. AHHM-FX-2020-02. 
\end{acknowledgments}

\appendix
\begin{widetext}
	\section{The standard method to calculate nonreciprocal free energy\label{standard}}
	This appendix provides detailed derivations of the free energy to first order in the magnetic field for both the Rashba and unconventional Rashba models, following the standard approach of Refs. \cite{RN81,RN98}. The superconductivity susceptibility in the coefficient $\alpha_{\boldsymbol{q}}$ is calculated by
	\begin{align}
		\chi(\boldsymbol{q})=k_BT\sum_{\boldsymbol{k},i\omega_n}\mathrm{Tr}\left[\Gamma^{\dagger}\mathcal{G}\left(\boldsymbol{k}+\frac{\boldsymbol{q}}{2},i\omega_n\right)\Gamma\mathcal{G}^T\left(-\boldsymbol{k}+\frac{\boldsymbol{q}}{2},-i\omega_n\right)\right]. \label{chi}
	\end{align}
	We use the Rashba system and unconventional Rashba system for examples. An in-plane magnetic field is applied to these models. To the first order of the field, the Green's function is
	\begin{align}
		\mathcal{G}\left(\boldsymbol{k},i\omega_n\right)=\left[i\omega_n-H_0(\boldsymbol{k})-\boldsymbol{B}\cdot\boldsymbol{\sigma}\right]^{-1}=\mathcal{G}_0\left(\boldsymbol{k},i\omega_n\right)+\mathcal{G}_0\left(\boldsymbol{k},i\omega_n\right)(\boldsymbol{B}\cdot\boldsymbol{\sigma})\mathcal{G}_0\left(\boldsymbol{k},i\omega_n\right), \label{gf}
	\end{align}
	where $\mathcal{G}_0\left(\boldsymbol{k},i\omega_n\right)$ is the bare Green's function without field. Plugging Eqs. (\ref{chi}) and (\ref{gf}), to the first order of the field, the susceptibility is contributed by two terms
	\begin{align}
		\chi(\boldsymbol{q})&=\chi_0(\boldsymbol{q})+\chi_{\boldsymbol{B}}(\boldsymbol{q}), \\
		\chi_0(\boldsymbol{q})&=k_BT\sum_{\boldsymbol{k},i\omega_n}\mathrm{Tr}\left[\Gamma^{\dagger}\mathcal{G}_0\left(\boldsymbol{k}+\frac{\boldsymbol{q}}{2},i\omega_n\right)\Gamma\mathcal{G}_0^T\left(-\boldsymbol{k}+\frac{\boldsymbol{q}}{2},-i\omega_n\right)\right], \label{chi0} \\
		\chi_{\boldsymbol{B}}(\boldsymbol{q})&=k_BT\sum_{\boldsymbol{k},i\omega_n}\left\{\mathrm{Tr}\left[\Gamma^{\dagger}\mathcal{G}_0\left(\boldsymbol{k}+\frac{\boldsymbol{q}}{2},i\omega_n\right)(\boldsymbol{B}\cdot\boldsymbol{\sigma})\mathcal{G}_0\left(\boldsymbol{k}+\frac{\boldsymbol{q}}{2},i\omega_n\right)\Gamma\mathcal{G}_0^T\left(-\boldsymbol{k}+\frac{\boldsymbol{q}}{2},-i\omega_n\right)\right]\right. \notag  \\
		&\left.+\mathrm{Tr}\left[\Gamma^{\dagger}\mathcal{G}_0\left(\boldsymbol{k}+\frac{\boldsymbol{q}}{2},i\omega_n\right)\Gamma\mathcal{G}_0^T\left(-\boldsymbol{k}+\frac{\boldsymbol{q}}{2},-i\omega_n\right)(\boldsymbol{B}\cdot\boldsymbol{\sigma}^T)\mathcal{G}_0^T\left(-\boldsymbol{k}+\frac{\boldsymbol{q}}{2},-i\omega_n\right)\right]\right\}. \label{chiB}
	\end{align}
	\subsection{Rashba system}
	For a Rashba system with Hamiltonian
	\begin{equation}
		H_R\left(\boldsymbol{k}\right)=\xi_{\boldsymbol{k}}\sigma^0+\lambda_{R}\left(\boldsymbol{k}\times\boldsymbol{\sigma}\right)\cdot\hat{\boldsymbol{z}},
	\end{equation}
	the Fermi surface can not be regarded as a single circle, but splits to inner and outer ones, we use $s\in\{+,-\}$ to label them. The two Fermi momenta $k_{F}^{s}=\left(\sqrt{\lambda_{R}^2+4t\mu}-s\lambda_{R}\right)/2t$, $s=+$ for the inner circle while $s=-$ for the outer circle, and $\xi_{\boldsymbol{k}}=tk^2-\mu$ is the standard electron dispersion. One can find the corresponding Fermi velocities are the same
	\begin{align}
		v_F\equiv2tk_F^{s}+s\lambda_{R}=\sqrt{\lambda_{R}^2+4t\mu}.
	\end{align}
	The Green's function for electron is
	\begin{align}
		\mathcal{G}_0\left(\boldsymbol{k},i\omega_n\right)=\left[i\omega_n-H_R(\boldsymbol{k})\right]^{-1}=g_{++}\sigma^0+g_{+-}\left(\hat{\boldsymbol{k}}\times\boldsymbol{\sigma}\right)\cdot\hat{\boldsymbol{z}},
	\end{align}
	with $g_{\alpha\beta}=\alpha/\left[i\omega_n-E_{+}(\boldsymbol{k})\right]+\beta/\left[i\omega_n-E_{-}(\boldsymbol{k})\right]$, and $E_{\pm}(\boldsymbol{k})=\xi_{\boldsymbol{k}}\pm\lambda_Rk$. Then we can calculate the superconductivity susceptibility with 
	\begin{align}
		\mathcal{G}_0\left(\boldsymbol{k}+\frac{\boldsymbol{q}}{2},i\omega_n\right)&=g_{++}\sigma^0+g_{+-}\left\{\left[\left(1-\frac{\hat{\boldsymbol{k}}\cdot\boldsymbol{q}}{2k}\right)\hat{\boldsymbol{k}}+\frac{\boldsymbol{q}}{2k}\right]\times\boldsymbol{\sigma}\right\}\cdot\hat{\boldsymbol{z}}, \label{G01} \\
		\mathcal{G}_0^T\left(-\boldsymbol{k}+\frac{\boldsymbol{q}}{2},-i\omega_n\right)&=h_{++}\sigma^0+h_{+-}\left\{\left[-\left(1+\frac{\hat{\boldsymbol{k}}\cdot\boldsymbol{q}}{2k}\right)\hat{\boldsymbol{k}}+\frac{\boldsymbol{q}}{2k}\right]\times\boldsymbol{\sigma}^T\right\}\cdot\hat{\boldsymbol{z}}. \label{G02}
	\end{align}
	Since the band splitting is much larger than the superconducting order parameter, we neglect the contribution from bands deviating from the Fermi surface (e.g., $g_{++}^+=g_{+-}^+=1/[i\omega_n-E_{+}(\boldsymbol{k})]$). There are the relations
	\begin{align}
		g_{++}^{s}\left(\boldsymbol{k},\boldsymbol{q},i\omega_n\right)&=s g_{+-}^{s}\left(\boldsymbol{k},\boldsymbol{q},i\omega_n\right), \\
		h_{++}^{s}\left(\boldsymbol{k},\boldsymbol{q},i\omega_n\right)&=s h_{+-}^{s}\left(\boldsymbol{k},\boldsymbol{q},i\omega_n\right).
	\end{align}
	And the explicit values to the order we care are
	\begin{align}
		&g_{++}^{s}=\frac{G}{2}\left[1+G\left(\boldsymbol{v}^{s}\cdot\frac{\boldsymbol{q}}{2}+s\lambda_{R}\left|\boldsymbol{k}+\frac{\boldsymbol{q}}{2}\right|\right)+G^2\left(\boldsymbol{v}^{s}\cdot\frac{\boldsymbol{q}}{2}\right)\left(\boldsymbol{v}^{s}\cdot\frac{\boldsymbol{q}}{2}+2s\lambda_{R}\left|\boldsymbol{k}+\frac{\boldsymbol{q}}{2}\right|\right) \notag\right. \\
		&\phantom{=\;\;}\left.+ G^3\left(\boldsymbol{v}^{s}\cdot\frac{\boldsymbol{q}}{2}\right)^2\left(\boldsymbol{v}^{s}\cdot\frac{\boldsymbol{q}}{2}+3s\lambda_{R}\left|\boldsymbol{k}+\frac{\boldsymbol{q}}{2}\right|\right)+ G^4\left(\boldsymbol{v}^{s}\cdot\frac{\boldsymbol{q}}{2}\right)^3\left(\boldsymbol{v}^{s}\cdot\frac{\boldsymbol{q}}{2}+4\nu\lambda_{R}\left|\boldsymbol{k}+\frac{\boldsymbol{q}}{2}\right|\right)\right], \label{g++} \\
		&h_{++}^{s}=-\frac{\tilde{G}}{2}\left[1+\tilde{G}\left(\boldsymbol{v}^{s}\cdot\frac{\boldsymbol{q}}{2}-s\lambda_{R}\left|-\boldsymbol{k}+\frac{\boldsymbol{q}}{2}\right|\right)+\tilde{G}^2\left(\boldsymbol{v}^{s}\cdot\frac{\boldsymbol{q}}{2}\right)\left(\boldsymbol{v}^{s}\cdot\frac{\boldsymbol{q}}{2}-2s\lambda_{R}\left|-\boldsymbol{k}+\frac{\boldsymbol{q}}{2}\right|\right) \notag\right. \\
		&\phantom{=\;\;}\left.+ \tilde{G}^3\left(\boldsymbol{v}^{s}\cdot\frac{\boldsymbol{q}}{2}\right)^2\left(\boldsymbol{v}^{s}\cdot\frac{\boldsymbol{q}}{2}-3s\lambda_{R}\left|-\boldsymbol{k}+\frac{\boldsymbol{q}}{2}\right|\right)+ \tilde{G}^4\left(\boldsymbol{v}^{s}\cdot\frac{\boldsymbol{q}}{2}\right)^3\left(\boldsymbol{v}^{s}\cdot\frac{\boldsymbol{q}}{2}-4s\lambda_{R}\left|-\boldsymbol{k}+\frac{\boldsymbol{q}}{2}\right|\right)\right]. \label{h++}
	\end{align}
	Here, the velocity $\boldsymbol{v}^{s}\equiv2t\boldsymbol{k}_F^{s}=\left(v_F-s\lambda_{R}\right)\hat{\boldsymbol{k}}$, the Green's functions are $G=(i\omega_n-\xi_{\boldsymbol{k}})^{-1}$ and $\tilde{G}=(i\omega_n+\xi_{-\boldsymbol{k}})^{-1}$. It is convenient to calculate the summation with the formula
	\begin{align}
		\langle &G^l\tilde{G}^m\rangle=k_BT\sum_{\boldsymbol{k},i\omega_n}\frac{1}{\left(i\omega_n-\xi_{\boldsymbol{k}}\right)^l}\frac{1}{\left(i\omega_n+\xi_{\boldsymbol{-k}}\right)^m}=\frac{N_F}{(l-1)!(m-1)!}\tilde{\chi}_{l+m}, \label{Chilm}\\
		&\tilde{\chi}_{l+m}=\begin{cases}
			-\ln\left(C/T\right),\quad l+m=2 \\
			\frac{4(l+m-2)!}{(2i)^{l+m}}\left[1+(-1)^{l+m}\right]\left[1-\left(\frac{1}{2}\right)^{l+m-1}\right]\frac{\zeta(l+m-1)}{\left(\pi k_BT\right)^{l+m-2}}. \label{chilm}
		\end{cases}
	\end{align}
	Here, $\zeta(x)$ is the Riemann zeta function. One can find that the nonzero contributions are the terms in even order of the Green's function, or to say, $l+m$ is even. Using the result
	\begin{align}
		&\left(gh\right)_{s}\equiv g_{++}^{s}h_{++}^{s}=g_{+-}^{s}h_{+-}^{s} \notag \\
		&=-\frac{G\tilde{G}}{4}\left\{1+\left(G^2+\tilde{G}^2+G\tilde{G}\right)\left(\boldsymbol{v}^{s}\cdot\frac{\boldsymbol{q}}{2}\right)^2+2s\left(\boldsymbol{v}^{s}\cdot\frac{\boldsymbol{q}}{2}\right)\lambda_{R}\left[\left(G^2+\frac{G\tilde{G}}{2}\right)\left|\boldsymbol{k}+\frac{\boldsymbol{q}}{2}\right|-\left(\tilde{G}^2+\frac{G\tilde{G}}{2}\right)\left|-\boldsymbol{k}+\frac{\boldsymbol{q}}{2}\right|\right]\right. \notag \\
		&+\left(G^4+\tilde{G}^4+G^3\tilde{G}+G\tilde{G}^3+G^2\tilde{G}^2\right)\left(\boldsymbol{v}^{s}\cdot\frac{\boldsymbol{q}}{2}\right)^4+s\left(\boldsymbol{v}^{s}\cdot\frac{\boldsymbol{q}}{2}\right)^3\lambda_{R}\left[\left(4G^4+3G^3\tilde{G}+G\tilde{G}^3+2G^2\tilde{G}^2\right)\left|\boldsymbol{k}+\frac{\boldsymbol{q}}{2}\right| \notag\right. \\
		&\phantom{=\;\;}\left.\phantom{=\;\;}\left.-\left(4\tilde{G}^4+G^3\tilde{G}+3G\tilde{G}^3+2G^2\tilde{G}^2\right)\left|-\boldsymbol{k}+\frac{\boldsymbol{q}}{2}\right|\right]\right\}, \label{gh}
	\end{align}
	thus,
	\begin{align}
		\left\langle\left(gh\right)_{s}\right\rangle=-\frac{N_F^{s}}{4}\left[\tilde{\chi}_2+\frac{1}{2}v_F^2\tilde{\chi}_4\left\langle\left(\hat{\boldsymbol{k}}\cdot\boldsymbol{q}\right)^2\right\rangle+\frac{1}{24}v_F^4\tilde{\chi}_6\left\langle\left(\hat{\boldsymbol{k}}\cdot\boldsymbol{q}\right)^4\right\rangle\right]. \label{gh}
	\end{align}
	The field free susceptibility is calculated as
	\begin{align}
		\chi_0(\boldsymbol{q})=2\sum_{s}\langle g_{++}^{s}h_{++}^{s}+g_{+-}^{s}h_{+-}^{s}\rangle=-\left(N_{F}^{+}+N_{F}^{-}\right)\left(\tilde{\chi}_2+\frac{1}{4}\tilde{\chi}_4v_F^2q^2\right)
	\end{align}
	The field-dependent susceptibility is contributed by values as follows
	\begin{align}
		\left(g^2h\right)_{s}&\equiv g_{++}^{s}g_{+-}^{s}h_{++}^{s}=\left(g_{++}^{s}\right)^2h_{+-}^{s}=\left(g_{+-}^{s}\right)^2h_{+-}^{s}, \\
		\left(gh^2\right)_{s}&\equiv g_{++}^{s}h_{++}^{s}h_{+-}^{s}=g_{+-}^{s}\left(h_{++}^{s}\right)^2=g_{+-}^{s}\left(h_{+-}^{s}\right)^2.
	\end{align}
	These values have nonzero parts which are proportional to $\hat{\boldsymbol{k}}\cdot\boldsymbol{q}$ and $\left(\hat{\boldsymbol{k}}\cdot\boldsymbol{q}\right)^3$. Therefore, we get the nonzero term of $\chi_{\boldsymbol{B}}$ 
	\begin{align}
		\chi_{\boldsymbol{B}}\left(\boldsymbol{q}\right)=8\sum_{s}\left\langle\left[\left(g^2h\right)_{s}-\left(gh^2\right)_{s}\right]\left(\hat{\boldsymbol{k}}\times\boldsymbol{B}\right)\cdot\hat{\boldsymbol{z}}\right\rangle. 
	\end{align}
	Calculations show that 
	\begin{align}
		\left\langle\left(g^2h\right)_{s}\left(\hat{\boldsymbol{k}}\times\boldsymbol{B}\right)\cdot\hat{\boldsymbol{z}}\right\rangle=-\left\langle\left(gh^2\right)_{s}\left(\hat{\boldsymbol{k}}\times\boldsymbol{B}\right)\cdot\hat{\boldsymbol{z}}\right\rangle=-\frac{N_F^{s}}{16}\left\{\left(s v^{s}+\lambda_{R}\right)\tilde{\chi}_4+\frac{1}{8}\left[s\left(v^{s}\right)^3+3\lambda_{R}\left(v^{s}\right)^2\right]\tilde{\chi}_6q^2\right\}\left(\boldsymbol{q}\times\boldsymbol{B}\right)\cdot\hat{\boldsymbol{z}}. 
	\end{align}
	Plugging $N_F^{s}=\left(1-s\lambda_{R}/v_F\right)N_F/2$, we get
	\begin{equation}
		\chi_{\boldsymbol{B}}\left(\boldsymbol{q}\right)=N_F\left(\tilde{\chi}_4+\frac{1}{8}\tilde{\chi}_6v_F^2q^2\right)\lambda_{R}\left(\boldsymbol{q}\times\boldsymbol{B}\right)\cdot\hat{\boldsymbol{z}}.
	\end{equation}
	The coefficient for the free energy is obtained
	\begin{align}
		\alpha_{\boldsymbol{q}}=\frac{1}{V}+N_F\left[\left(\tilde{\chi}_2+\frac{1}{4}\tilde{\chi}_4v_F^2q^2\right)-\left(\tilde{\chi}_4+\frac{1}{8}\tilde{\chi}_6v_F^2q^2\right)\lambda_{R}\boldsymbol{q}\cdot\left(\boldsymbol{B}\times\hat{\boldsymbol{z}}\right)\right]. \label{aq}
	\end{align}
	\subsection{Unconventional Rashba system}
	The bare Green's function for unconventional Rashba model is
	\begin{align}
		\mathcal{G}_0(\boldsymbol{k},i\omega_n)&=G_{++++}\sigma^0\tau^0+G_{+-+-}\hat{\boldsymbol{k}}\cdot\left(\boldsymbol{\sigma}\times\hat{\boldsymbol{z}}\right)\tau^x+G_{++--}\left[\frac{\lambda_Rk}{\sqrt{\lambda^2+\lambda_R^2k^2}}\hat{\boldsymbol{k}}\cdot\left(\boldsymbol{\sigma}\times\hat{\boldsymbol{z}}\right)\tau^0+\frac{\lambda}{\sqrt{\lambda^2+\lambda_R^2k^2}}\sigma^z\tau^y\right] \notag \\
		&+G_{+--+}\left[\frac{\lambda_Rk}{\sqrt{\lambda^2+\lambda_R^2k^2}}\sigma^0\tau^x-\frac{\lambda}{\sqrt{\lambda^2+\lambda_R^2k^2}}\hat{\boldsymbol{k}}\cdot\boldsymbol{\sigma}\tau^z\right], 
	\end{align}
	with $G_{\alpha\beta\gamma\delta}=(\alpha G_{++}+\beta G_{+-}+\gamma G_{-+}+\delta G_{--})/4$, and 
	\begin{align}
		&G_{\alpha\beta}(\boldsymbol{k},i\omega_n)=\frac{1}{i\omega_n-E_{\alpha\beta}(\boldsymbol{k})}, \\
		&E_{\alpha\beta}(\boldsymbol{k})=\xi_{\boldsymbol{k}}+\alpha\sqrt{\lambda^{2}+\lambda_{R}^{2}k^{2}}+\beta\varepsilon\lambda_{R}k.
	\end{align}
	Since the Fermi surface is setted in the gap, we ignore the contribution from the upper two bands. We can simplify the Green's function to 
	\begin{align}
		&\mathcal{G}_0(\boldsymbol{k},i\omega_n)=G_{++++}\left[\sigma^0\tau^0-\frac{\lambda_Rk}{\sqrt{\lambda^2+\lambda_R^2k^2}}\hat{\boldsymbol{k}}\cdot\left(\boldsymbol{\sigma}\times\hat{\boldsymbol{z}}\right)\tau^0-\frac{\lambda}{\sqrt{\lambda^2+\lambda_R^2k^2}}\sigma^z\tau^y\right] \notag \\
		&+G_{+-+-}\left[\hat{\boldsymbol{k}}\cdot\left(\boldsymbol{\sigma}\times\hat{\boldsymbol{z}}\right)\tau^x-\frac{\lambda_Rk}{\sqrt{\lambda^2+\lambda_R^2k^2}}\sigma^0\tau^x+\frac{\lambda}{\sqrt{\lambda^2+\lambda_R^2k^2}}\hat{\boldsymbol{k}}\cdot\boldsymbol{\sigma}\tau^z\right].
	\end{align}
	Ignoring pairings deviating from Fermi surface, for electron and hole, we get the Green's functions
	\begin{align}
		&\mathcal{G}_0(\boldsymbol{k}+\frac{\boldsymbol{q}}{2},i\omega_n)=\frac{\tilde{g}_{++}}{2}\left[\sigma^0\tau^0-\sin\phi_{\boldsymbol{q}}\frac{\boldsymbol{k}+\boldsymbol{q}/2}{\left|\boldsymbol{k}+\boldsymbol{q}/2\right|}\cdot\left(\boldsymbol{\sigma}\times\hat{\boldsymbol{z}}\right)\tau^0-\cos\phi_{\boldsymbol{q}}\sigma^z\tau^y\right] \notag \\
		&+\frac{\tilde{g}_{+-}}{2}\left[\frac{\boldsymbol{k}+\boldsymbol{q}/2}{\left|\boldsymbol{k}+\boldsymbol{q}/2\right|}\cdot\left(\boldsymbol{\sigma}\times\hat{\boldsymbol{z}}\right)\tau^x-\sin\phi_{\boldsymbol{q}}\sigma^0\tau^x+\cos\phi_{\boldsymbol{q}}\frac{\boldsymbol{k}+\boldsymbol{q}/2}{\left|\boldsymbol{k}+\boldsymbol{q}/2\right|}\cdot\boldsymbol{\sigma}\tau^z\right], \label{URG1} \\
		&\mathcal{G}_0^T(-\boldsymbol{k}+\frac{\boldsymbol{q}}{2},-i\omega_n)=\frac{\tilde{h}_{++}}{2}\left[\sigma^0\tau^0+\sin\phi_{-\boldsymbol{q}}\frac{\boldsymbol{k}-\boldsymbol{q}/2}{\left|\boldsymbol{k}-\boldsymbol{q}/2\right|}\cdot\left(\boldsymbol{\sigma}^T\times\hat{\boldsymbol{z}}\right)\tau^0+\cos\phi_{-\boldsymbol{q}}\sigma^z\tau^y\right] \notag \\
		&-\frac{\tilde{h}_{+-}}{2}\left[\frac{\boldsymbol{k}-\boldsymbol{q}/2}{\left|\boldsymbol{k}-\boldsymbol{q}/2\right|}\cdot\left(\boldsymbol{\sigma}^T\times\hat{\boldsymbol{z}}\right)\tau^x+\sin\phi_{-\boldsymbol{q}}\sigma^0\tau^x+\cos\phi_{-\boldsymbol{q}}\frac{\boldsymbol{k}-\boldsymbol{q}/2}{\left|\boldsymbol{k}-\boldsymbol{q}/2\right|}\cdot\boldsymbol{\sigma}^T\tau^z\right]. \label{URG2}
	\end{align}
	Here, the Green's functions correspond to the previous values Eqs. (\ref{g++},\ref{h++}) by $\tilde{g}(\tilde{h})=g\left(h\right)|_{\lambda_{R}\to\varepsilon\lambda_{R}}$. Then we get the susceptibilities
	\begin{align}
		\chi_0\left(\boldsymbol{q}\right)&=2\sum_{s}\left\langle \left(1+\cos\phi_{\boldsymbol{q}}\cos\phi_{-\boldsymbol{q}}+\sin\phi_{\boldsymbol{q}}\sin\phi_{-\boldsymbol{q}}\right)\left(gh\right)_{s}\right\rangle=-N_F\left(\tilde{\chi}_2+\frac{1}{4}\tilde{\chi}_4v_F^2q^2\right). 
	\end{align}
	We have ignored the value to the order of $\left(q/k_F\right)^2$. The field-dependent susceptibility is related to the values
	\begin{align}
		&\left(g_{++}^{s}\right)^2h_{++}^{s}=\left(g_{+-}^{\nu}\right)^2h_{++}^{s}=g_{++}^{s}g_{+-}^{s}h_{+-}^{s}=s\left(g^2h\right)_{s}, \\
		&g_{++}^{s}\left(h_{++}^{s}\right)^2=g_{++}^{s}\left(h_{+-}^{s}\right)^2=g_{+-}^{s}h_{++}^{s}h_{+-}^{s}=s\left(gh^2\right)_{s}.
	\end{align}
	And the result is
	\begin{align}
		\chi_{\boldsymbol{B}}\left(\boldsymbol{q}\right)&=-8\sum_{s}s\sin\phi\left\langle\left[\left(g^2h\right)_{s}-\left(gh^2\right)_{s}\right]\left(\hat{\boldsymbol{k}}\times\boldsymbol{B}\right)\cdot\hat{\boldsymbol{z}}\right\rangle \notag \\
		&=N_F\left(\tilde{\chi}_4+\frac{1}{8}\tilde{\chi}_6v_F^2q^2\right)\frac{\lambda_Rk_F}{\sqrt{\lambda^2+\lambda_R^2k_F^2}}v_F\left(\boldsymbol{q}\times\boldsymbol{B}\right)\cdot\hat{\boldsymbol{z}} \notag \\
		&\approx\frac{2\mu}{\lambda}N_F\left(\tilde{\chi}_4+\frac{1}{8}\tilde{\chi}_6v_F^2q^2\right)\lambda_R\left(\boldsymbol{q}\times\boldsymbol{B}\right)\cdot\hat{\boldsymbol{z}}. 
	\end{align}
	With the assumption of $\lambda\approx\mu$, we get the final line. This result shows that $\chi_{\boldsymbol{B}}^{UR}=\left(2\mu/\lambda\right)\chi_{\boldsymbol{B}}^{R}$. 
	
	\section{The simplified method to calculate nonreciprocal free energy\label{simplified}}
	We simplify the calculation by writting the superconducting model in band space, for example, the Hamiltonian for Rashba system can be written as
	\begin{align}
		H_R=\sum_{\boldsymbol{k},s}\xi_{\boldsymbol{k}s}c^{\dagger}_{\boldsymbol{k}s}c_{\boldsymbol{k}s}-V\sum_{\boldsymbol{k},\boldsymbol{k'},s}c^{\dagger}_{\boldsymbol{k}s}c^{\dagger}_{-\boldsymbol{k}s}c_{-\boldsymbol{k'}s}c_{\boldsymbol{k'}s},
	\end{align}
	where $s\in\{+,-\}$ denotes the chiral spin configuration in two types. Without external field, the eigenstates for free electrons are $\xi_{\boldsymbol{k}s}=tk^2-\mu+s\lambda_Rk$. Applying an in-plane magnetic field, the system is discribed by a mean field model
	\begin{align}
		H_R(\boldsymbol{B})=\sum_{\boldsymbol{k},s}\left[\tilde{\xi}_{\boldsymbol{k}s}c^{\dagger}_{\boldsymbol{k},s}c_{\boldsymbol{k},s}+\Delta(\boldsymbol{q})c^{\dagger}_{\boldsymbol{k}+\frac{\boldsymbol{q}}{2},s}c^{\dagger}_{-\boldsymbol{k}+\frac{\boldsymbol{q}}{2},s}+\Delta^*(\boldsymbol{q})c_{-\boldsymbol{k}+\frac{\boldsymbol{q}}{2},s}c_{\boldsymbol{k}+\frac{\boldsymbol{q}}{2},s}\right]+\frac{|\Delta(\boldsymbol{q})|^2}{V}. 
	\end{align}
	The eigenenergy for free electron is $\tilde{\xi}_{\boldsymbol{k}s}(\boldsymbol{B})=tk^2-\mu+s\left|\lambda_R\boldsymbol{k}+\boldsymbol{B}\times\hat{\boldsymbol{z}}\right|$, which is approximated to $\tilde{\xi}_{\boldsymbol{k}s}(\boldsymbol{B})=tk^2-\mu+s\left[\lambda_Rk+\hat{\boldsymbol{k}}\cdot\left(\boldsymbol{B}\times\hat{\boldsymbol{z}}\right)\right]$ in the weak field limit. The coefficient for $|\Delta(\boldsymbol{q})|^2$ free energy term is calculated as
	\begin{equation}
		\alpha_{\boldsymbol{q}}=\frac{1}{V}-k_BT\sum_{\boldsymbol{k},i\omega_n,s}\mathcal{G}_s(\boldsymbol{k}+\frac{\boldsymbol{q}}{2},i\omega_n)\mathcal{G}_s(-\boldsymbol{k}+\frac{\boldsymbol{q}}{2},-i\omega_n). \label{alphaq}
	\end{equation}
	Here, $\mathcal{G}_s(\boldsymbol{k},i\omega_n)=\left(i\omega_n-\tilde{\xi}_{\boldsymbol{k}s}\right)^{-1}$ is the electron's Matsubara Green's function, which can be expanded to the series of the field and momentum $\boldsymbol{q}$. To the first order, the eigenenergy is
	\begin{align}
		\tilde{\xi}_{\nu\boldsymbol{k}+\frac{\boldsymbol{q}}{2},s}=\xi_{\boldsymbol{k}s}+\nu\frac{\boldsymbol{q}}{2}\cdot\nabla_{\boldsymbol{k}}\xi_{\boldsymbol{k}s}+s\nu\left(1+\frac{\boldsymbol{q}}{2}\cdot\nabla_{\boldsymbol{k}}\right)\hat{\boldsymbol{k}}\cdot\left(\boldsymbol{B}\times\hat{\boldsymbol{z}}\right)=\xi_{\boldsymbol{k}s}+\delta_{\boldsymbol{k}s}^{\nu}. \label{dks}
	\end{align}
	We assume the small value of $\delta_{\boldsymbol{k}s}/k_BT$, thus the Matsubara Green's function can be expanded as
	\begin{align}
		\mathcal{G}_s(\nu\boldsymbol{k}+\frac{\boldsymbol{q}}{2},\nu i\omega_n)=\mathcal{G}_{\nu\boldsymbol{k}s}^0+\sum_{l=1}^{\infty}\left(\mathcal{G}_{\nu\boldsymbol{k}s}^0\right)^{l+1}\left(\delta_{\boldsymbol{k}s}^{\nu}\right)^l.
	\end{align}
	The Green's function $\mathcal{G}_{\nu\boldsymbol{k}s}^0=\left(\nu i\omega_n-\xi_{\boldsymbol{k}s}\right)^{-1}$ is the zero momentum one for band $s$ without field. We define a function for convenience
	\begin{align}
		G_{l,m}=k_BT\sum_{\boldsymbol{k},i\omega_n}\left(\mathcal{G}_{\boldsymbol{k}s}^0\right)^l\left(\mathcal{G}_{-\boldsymbol{k}s}^0\right)^m=\frac{(-1)^{m}N_{F,s}}{(l-1)!(m-1)!}\tilde{\chi}_{l+m},
	\end{align}
	where $\tilde{\chi}_{l+m}$ is in the same defination as Eq. (\ref{chilm}), and $N_{F,s}$ is the density of states on band $s$. Therefore the summation of the product in Eq. (\ref{alphaq}) is
	\begin{align}
		k_BT\sum_{\boldsymbol{k},i\omega_n}\mathcal{G}_s(\boldsymbol{k}+\frac{\boldsymbol{q}}{2},i\omega_n)\mathcal{G}_s(-\boldsymbol{k}+\frac{\boldsymbol{q}}{2},-i\omega_n)=G_{1,1}+\sum_{l=1}^{\infty}\left[G_{1,l+1}\left(\delta_{\boldsymbol{k}s}^{-}\right)^l+G_{l+1,1}\left(\delta_{\boldsymbol{k}s}^{+}\right)^l\right]+\sum_{l,m}G_{l+1,m+1}\left(\delta_{\boldsymbol{k}s}^{+}\right)^l\left(\delta_{\boldsymbol{k}s}^{-}\right)^m. \label{sum1}
	\end{align}
	We shall see, once the $\delta_{\boldsymbol{k}s}$ is obtained, the free energy will be solved. 
	\begin{align}
		\delta_{\boldsymbol{k}s}^{\nu}&=\nu\left[v_{F,s}\hat{\boldsymbol{k}}\cdot\frac{\boldsymbol{q}}{2}+s\left(\hat{\boldsymbol{k}}+\frac{\boldsymbol{q}}{2k}-\frac{\hat{\boldsymbol{k}}\cdot\boldsymbol{q}}{2k}\hat{\boldsymbol{k}}\right)\cdot\left(\boldsymbol{B}\times\hat{\boldsymbol{z}}\right)\right], \\
		\left\langle\delta_{\boldsymbol{k}s}^2\right\rangle_{\boldsymbol{k}}&=\frac{1}{8}v_F^2q^2+\frac{1}{2}sv_F\boldsymbol{q}\cdot\left(\boldsymbol{B}\times\hat{\boldsymbol{z}}\right), \label{d2}\\
		\left\langle\delta_{\boldsymbol{k}s}^4\right\rangle_{\boldsymbol{k}}&=\frac{3}{16}sv_F^3q^2\boldsymbol{q}\cdot\left(\boldsymbol{B}\times\hat{\boldsymbol{z}}\right). \label{d4}
	\end{align}
	Cutting the result to the fourth order of $\delta_{\boldsymbol{k}s}/k_BT$, we get the value
	\begin{align}
		&G_{1,1}+\left[G_{2,2}\delta_{\boldsymbol{k}s}^{+}\delta_{\boldsymbol{k}s}^{-}+G_{1,3}\left(\delta_{\boldsymbol{k}s}^{-}\right)^2+G_{3,1}\left(\delta_{\boldsymbol{k}s}^{+}\right)^2\right] \notag \\
		&+\left[G_{3,3}\left(\delta_{\boldsymbol{k}s}^{+}\right)^2\left(\delta_{\boldsymbol{k}s}^{-}\right)^2+G_{2,4}\delta_{\boldsymbol{k}s}^{+}\left(\delta_{\boldsymbol{k}s}^{-}\right)^3+G_{4,2}\left(\delta_{\boldsymbol{k}s}^{+}\right)^3\delta_{\boldsymbol{k}s}^{-}+G_{1,5}\left(\delta_{\boldsymbol{k}s}^{-}\right)^4+G_{5,1}\left(\delta_{\boldsymbol{k}s}^{+}\right)^4\right] \notag \\
		=&-N_{F,s}\left(\tilde{\chi}_2+2\tilde{\chi}_4\delta_{\boldsymbol{k}s}^2+\frac{2}{3}\tilde{\chi}_6\delta_{\boldsymbol{k}s}^4\right). \label{sum2}
	\end{align}
	Plugging $N_{F,s}=(N_F/2)(1-s\lambda_R/v_F)$ and Eqs. (\ref{alphaq},\ref{sum1},\ref{d2},\ref{d4},\ref{sum2}), the coefficient Eq. (\ref{aq}) is obtained again. For unconventional Rashba systems, the eigenenergy for free electron changes to
	\begin{equation}
		\tilde{\xi}_{\boldsymbol{k}s}'=\underbrace{\xi_{\boldsymbol{k}s}-\sqrt{\lambda^2+\lambda_R^2k^2}}_{\xi_{\boldsymbol{k}s}'}-\frac{\lambda_R\boldsymbol{k}}{\sqrt{\lambda^2+\lambda_R^2k^2}}\cdot\left(\boldsymbol{B}\times\hat{\boldsymbol{z}}\right).
	\end{equation}
	Eq. (\ref{dks}) is substituted by
	\begin{align}
		\tilde{\xi}'_{\nu\boldsymbol{k}+\frac{\boldsymbol{q}}{2},s}=\xi_{\boldsymbol{k}s}'+\nu\frac{\boldsymbol{q}}{2}\cdot\nabla_{\boldsymbol{k}}\xi_{\boldsymbol{k}s}'-\nu\left(1+\frac{\boldsymbol{q}}{2}\cdot\nabla_{\boldsymbol{k}}\right)\frac{\lambda_R\boldsymbol{k}}{\sqrt{\lambda^2+\lambda_R^2k^2}}\cdot\left(\boldsymbol{B}\times\hat{\boldsymbol{z}}\right)=\xi_{\boldsymbol{k}s}'+\nu\delta_{\boldsymbol{k}s},
	\end{align}
	where 
	\begin{align}
		\delta_{\boldsymbol{k}s}=v_{F,s}\hat{\boldsymbol{k}}\cdot\frac{\boldsymbol{q}}{2}-\frac{\lambda_R}{\sqrt{\lambda^2+\lambda_R^2k^2}}\left(\boldsymbol{k}+\frac{\boldsymbol{q}}{2}\right)\cdot\left(\boldsymbol{B}\times\hat{\boldsymbol{z}}\right).
	\end{align}
	Repeat the calculation, we get
	\begin{align}
		\alpha_{\boldsymbol{q}}=\frac{1}{V}+N_F\left(\tilde{\chi}_2+\frac{1}{4}\tilde{\chi}_4v_F^2q^2\right)-N_F\left(\tilde{\chi}_4+\frac{1}{8}\tilde{\chi}_6v_F^2q^2\right)\frac{\lambda_Rk}{\sqrt{\lambda^2+\lambda_R^2k^2}}v_F\boldsymbol{q}\cdot\left(\boldsymbol{B}\times\hat{\boldsymbol{z}}\right).
	\end{align}
	
	Notably, the band energy difference between states with momenta $\boldsymbol{k}$ and $-\boldsymbol{k}$ can be expressed as $\Delta E_s(\boldsymbol{k})=(\delta_{\boldsymbol{k}s}^+-\delta_{\boldsymbol{k}s}^-)|_{\boldsymbol{q}=\boldsymbol{0}}$. Therefore, $\delta_{\boldsymbol{k}s}^{\nu}=\nu\delta_{\boldsymbol{k}s}$ decomposes into the $\boldsymbol{q}$-dependent and -independent terms as
	\begin{equation}
		\delta_{\boldsymbol{k}s}=\frac{1}{2}\left[\boldsymbol{v}_{F,s}\cdot\boldsymbol{q}+\Delta E_s(\boldsymbol{k})\right]. 
	\end{equation}
	Ignoring the factor from averaging the Fermi surface, the contribution to nonreciprocity is obtained by even powers of $\delta_{\boldsymbol{k}s}$ and proportional to $\sum_{s,l}q^{2l+1}C_{l}N_{F,s}[v_{F,s}(k_{F,s})]^{2l+1} \Delta E_s$, where $l$ are integers and $C_{l}$ are $l$-dependent factors. Plugging $N_{F,s}\propto k_{F,s}/v_{F,s}$, we get the nonreciprocal contribution as
	\begin{equation}
		\sum_{s,l}q^{2l+1}C_{l}k_{F,s}\left[v_{F,s}(k_{F,s})\right]^{2l} \Delta E_s. 
	\end{equation}
	
	\section{The theoretical-limit $\alpha _{\boldsymbol{q}}$ for Rashba system\label{fake}}
	Starting with Eqs. (\ref{chi0},\ref{chiB},\ref{G01},\ref{G02}), we get Green's functions different from the ones in the standard method, since we do not ignore the contribution from the band deviating from the Fermi surface. They are
	\begin{align}
		g_{++}\left(\boldsymbol{k},\boldsymbol{q},i\omega_n\right)&=G+\left(\boldsymbol{v_F}\cdot\boldsymbol{\frac{q}{2}}\right)G^2+\left(\boldsymbol{v_F}\cdot\boldsymbol{\frac{q}{2}}\right)^2G^3, \label{g2} \\
		h_{++}\left(\boldsymbol{k},\boldsymbol{q},i\omega_n\right)&=-\tilde{G}-\left(\boldsymbol{v_F}\cdot\boldsymbol{\frac{q}{2}}\right)\tilde{G}^2-\left(\boldsymbol{v_F}\cdot\boldsymbol{\frac{q}{2}}\right)^2\tilde{G}^3, \\
		g_{+-}\left(\boldsymbol{k},\boldsymbol{q},i\omega_n\right)&=\lambda_RkG^2+2\lambda_Rk\left(\boldsymbol{v_F}\cdot\boldsymbol{\frac{q}{2}}\right)G^3+3\lambda_Rk\left(\boldsymbol{v_F}\cdot\boldsymbol{\frac{q}{2}}\right)^2G^4, \\
		h_{+-}\left(\boldsymbol{k},\boldsymbol{q},i\omega_n\right)&=\lambda_Rk\tilde{G}^2+2\lambda_Rk\left(\boldsymbol{v_F}\cdot\boldsymbol{\frac{q}{2}}\right)\tilde{G}^3+3\lambda_Rk\left(\boldsymbol{v_F}\cdot\boldsymbol{\frac{q}{2}}\right)^2\tilde{G}^4. \label{h2}
	\end{align}
	Here, Green's functions are still $G=(i\omega_n-\xi_{\boldsymbol{k}})^{-1}$ and $\tilde{G}=(i\omega_n+\xi_{-\boldsymbol{k}})^{-1}$, the Fermi velocity $\boldsymbol{v_F}=\nabla_{\boldsymbol{k}} \xi_{\boldsymbol{k}}$. Plugging (\ref{chi0},\ref{chiB},\ref{G01},\ref{G02}), the field-independent and -dependent susceptibilities are gotten
	\begin{align}
		\chi_0(\boldsymbol{q})&=2\langle g_{++}h_{++}\rangle=-N_F\left(\tilde{\chi}_2+\frac{1}{4}\tilde{\chi}_4v_F^2q^2\right), \\
		\chi_{\boldsymbol{B}}(\boldsymbol{q})&=\langle2g_{++}h_{++}\left(g_{+-}+h_{+-}\right)-g_{++}^2h_{+-}-h_{++}^2g_{+-}\rangle\left(\frac{\boldsymbol{q}}{k}\times\boldsymbol{B}\right)\cdot\hat{\boldsymbol{z}}=-2N_{F}\left(\tilde{\chi}_4+\frac{1}{4}\tilde{\chi}_6v_F^2q^2\right)\lambda_R\left(\boldsymbol{q}\times\boldsymbol{B}\right)\cdot\hat{\boldsymbol{z}}.
	\end{align}
	The theoretical-limit "$\alpha_{\boldsymbol{q}}$" is
	\begin{align}
		\alpha_{\boldsymbol{q}}=\frac{1}{V}+N_F\left[\tilde{\chi}_2+\frac{1}{4}\tilde{\chi}_4v_F^2q^2+2\left(\tilde{\chi}_4+\frac{1}{4}\tilde{\chi}_6v_F^2q^2\right)\lambda_R\left(\boldsymbol{q}\times\boldsymbol{B}\right)\cdot\hat{\boldsymbol{z}}\right]. 
	\end{align}
	This result has nonreciprocity with a minus sign comparing with the real one, and it also gives the wrong quantitative value. However, it is sufficient to reveal the band asymmetry from this theoretical-limit "$\alpha _{\boldsymbol{q}}$". 
\end{widetext}


\begin{thebibliography}{55}
\bibitem [{not()}]{note}%
	\BibitemOpen
	\bibfield  {title} {\bibinfo {title} {For example, a model with multi-{DOF}
			has a part categorized to symmetry $\{T\zeta,I\zeta'\}$ explicitly written as
			${H}^{T\zeta,I\zeta'}=f_0(\boldsymbol{k})\boldsymbol{1}+f_1(\boldsymbol{k})\kappa_1^z+f_2(\boldsymbol{k})\kappa_2^z+f_3(\boldsymbol{k})\kappa_3^z$,
			where $\kappa_1^z$, $\kappa_2^z$ and $\kappa_3^z$ are pauli matrices for
			different {DOF}. {W}e need $n_{\zeta\zeta'}=2$ to generate the product
			matrices: $\boldsymbol{1}$, $\kappa_1^z$, $\kappa_2^z$, $\kappa_3^z$,
			$\kappa_1^z\kappa_2^z$, $\kappa_1^z\kappa_3^z$, $\kappa_2^z\kappa_3^z$ and
			$\kappa_1^z\kappa_2^z\kappa_3^z$}}\href@noop {} {\ }\BibitemShut
	{NoStop}%
\end{thebibliography}
\end{document}